\documentclass[%
reprint,
superscriptaddress,
groupedaddress,
nofootinbib,
 amsmath,amssymb,showpacs,
 aps,
 include graphics,
]{revtex4-2}
\usepackage{graphicx}
\usepackage{dcolumn}
\usepackage{bm}
\usepackage{float}
\usepackage[caption=false]{subfig}
\usepackage{tabularx}
\usepackage{multirow}  
\usepackage{xcolor}
\usepackage{placeins}
\usepackage{slashed}
\usepackage{url}
\usepackage{xcolor}
\definecolor{mycolor}{RGB}{0,0,204}
\usepackage{hyperref}
\hypersetup{
  colorlinks   = true,    
  urlcolor     = blue,    
  linkcolor    =blue,    
  citecolor    = blue      
}

\begin{document}


\title{\boldmath The Stochastic Gravitational-Wave Background from Primordial Black Holes in R-Symmetric $SU(5)$ Inflation
}

\author{Nadir Ijaz}
 \email{nadirijaz1919@gmail.com }
 
\author{Maria Mehmood}
 \email{mehmood.maria786@gmail.com}

\author{Mansoor Ur Rehman}
 \email{mansoor@qau.edu.pk}
\affiliation{Department of Physics, Quaid-i-Azam University, Islamabad 45320, Pakistan
}

\date{\today}

\begin{abstract}
This study explores the realization of nonminimally coupled Higgs inflation in the context of no-scale supergravity, investigates the formation of primordial black holes, and examines the potential for observable proton decay within the framework of the R-symmetric $SU(5)$ model. For inflation, both single and multifield scenarios are investigated. The prediction of the single-field model for the tensor-to-scalar ratio, $r$, is approximately $10^{-3}$, and the scalar spectral index falls within Planck’s 1$\sigma$ range. The running of the scalar spectral index, $-{dn_{s}}/{d\ln{k}}$, is approximately $10^{-4}$. A realistic scenario of reheating and non-thermal leptogenesis is employed with reheat temperature $T_r\sim10^9$ GeV. In the multifield case, we mainly focus on Primordial Black Holes (PBHs) and Gravitational Waves (GWs). In this inflationary framework, we demonstrate how a suitable choice of parameters can result in an enhanced scalar power spectrum, leading to the production of primordial black holes (PBHs) capable of fully accounting for dark matter.
We also show that this scenario leads to Scalar Induced Gravitational Waves (SIGW) which can be detected in current and future GW detectors. We explore different proton decay channels to look for observable predictions for the next-generation proton decay experiments Hyper-K and DUNE consistent with gauge coupling unification and cosmological bounds.
\end{abstract}

\maketitle


\section{\label{sec:level1}Introduction}
Over half a century ago, scientists first proposed the existence of primordial black holes (PBHs), which continue to captivate as intriguing theoretical phenomena \cite{Hawking2,hawking1974black}. In recent years, researchers have recognized the compelling potential of PBHs in explaining dark matter. Rather than necessitating the existence of yet-to-be-discovered elementary particles beyond the Standard Model, it is conceivable that dark matter could be comprised of a significant abundance of PBHs that emerged during the early stages of cosmic evolution.

During the radiation-dominated era of the early Universe, if the density contrast is sufficiently significant, the mass encompassed by a Hubble volume can undergo gravitational collapse, giving rise to the formation of black holes. These PBHs can span a much broader range of masses compared to black holes formed through the conventional process of stellar evolution. For instance, a population of PBHs with $M_{\text{pbh}} \sim \mathcal{O}(10)M_{\odot}$ could potentially explain the observed binary black hole merger events detected by LIGO-Virgo \cite{PhysRevLett.116.241103,PhysRevLett.116.061102,PhysRevLett.119.141101}. Additionally, PBHs with $M_{\text{pbh}} \sim \mathcal{O}(10^{5})M_{\odot}$ could have acted as seeds for the formation of supermassive black holes found at the centers of galaxies. However, if PBHs are considered to account for the entirety of the observed dark matter abundance, various theoretical and observational constraints impose restrictions on the mass range of PBHs, which would lie in a much lighter regime. Specifically, the allowed mass range for PBHs to explain dark matter would be between $10^{-17}M_{\odot}$ and $10^{-12}M_{\odot}$ \cite{Montero_Camacho_2019,Carr_2021}. Besides the formation of primordial black holes~\cite{Carr:1993aq,Kawasaki:1997ju,Yokoyama:1998pt,Kawasaki:1998vx,Khlopov:2008qy,Carr:2009jm,Lyth:2011kj,Drees:2011yz,Drees:2011hb,Ballesteros:2017fsr,Pi:2017gih,Ballesteros:2022hjk}, the amplified perturbations in scalar curvature are anticipated to generate significant secondary tensor perturbations. This has been explored in recent studies~ \cite{Di_2018,Bartolo_2019,Hajkarim_2020,Liu_2020,clesse2018detecting,ghoshal2023cosmological,Balaji:2023ehk,Cai:2023uhc,Basilakos:2023xof,Ijaz:2024zma,Conaci:2024tlc,Kawai:2021edk}, resulting in detectable gravitational wave signals in the upcoming detectors. See Refs.~\cite{Domenech:2021ztg,LISACosmologyWorkingGroup:2023njw} for recent reviews on SIGW.

We have studied Higgs inflation in the R-symmetric $SU(5)$ model based on a no-scale supergravity framework. A multifield treatment is required for the formation of primordial black holes which can also explain dark matter in totality.
In $SU(5)$ models, we often encounter a challenge known as the doublet-triplet splitting problem, which can be resolved by using the missing doublet mechanism. It has been studied in the literature using the GUT Higgs of the $24$ representation in \cite {Kounnas:1983gm,Berezhiani:1996nu,Antusch:2014poa,Antusch:2023mxx}, and utilizing the GUT Higgs of the $75$ representation in \cite{Masiero:1982fe,Hisano:1994fn,Hisano:1997nu,Berezhiani:1996nu,Altarelli:2000fu,Antusch:2014poa, Bajc:2016qcc,Pokorski:2019ete,Ellis:2021fhb,Mehmood:2023gmm}.
In this paper, we have employed the missing partner mechanism with GUT Higgs in the $24$-plet representation using only one pair of $50$-plets along with the minimal content of $SU(5)$.  The inclusion of a pair of $50$-plets provides heavy masses to color triplets while leaving $SU(2)_L$ doublets in the $5$-plet Higgs massless. 
We systematically investigated the parameter space to assess the potential for observable proton decay, specifically focusing on various decay channels mediated by color triplets. Additionally, we conducted an analysis of gauge coupling unification, taking into account the presence of color octets and $SU(2)_L$ triplets at an intermediate scale.

Section \ref{Mod} outlines the structure of the $SU(5)$ model employed in this study. In section \ref{sec:level2}, we explore Higgs inflation in the R-symmetric $SU(5)$ model. Moving forward to section \ref{PBHGW}, we investigate the presence of PBHs and gravitational waves within our model. Section \ref{gcusec} is dedicated to our study of two-loop Renormalization Group Equations (RGEs) for gauge couplings and matching conditions for the Minimal Supersymmetric Standard Model (MSSM) and R-symmetric $SU(5)$ model. In section \ref{pd}, we explored proton decay predictions and parameter range to look for observable proton decay in next-generation experiments. Finally, in section \ref{con}, we concluded our findings.
\section{\boldmath R-symmetric $SU(5)$-Missing Doublet Model}\label{Mod}
\begin{table}[t!]
\caption{\label{sf} The superfield content of $SU(5)$ and charge assignments under MSSM.}
\begin{ruledtabular}
\begin{tabular}{ll}
\textrm{~~$SU(5)$}&
\textrm{$SU(3)_c\times SU(2)_L \times U(1)_Y$}\\
\hline
 $~~~10_i$  & $Q_i ( 3, 2, 1/6)$
 + $U^c_i ( \overline{3}, 1, -2/3)$  
 + $E^c_i ( 1, 1, 1)$ \\
 \hline
$~~~\overline{5}_i$ & $D^c_i ( \overline{3}, 1, 1/3)$
+ $L_i ( 1, 2, -1/2)$\\
\hline
$~~~1_i$ & $N^c_i ( 1, 1, 0)$  \\
\hline
 $~~~5_{h}$  & $H_{u}(1,2,+1/2)$ 
 + $H_{T}(3,1,-1/3)$ \\
 \hline 
$~~~\bar{5}_{h}$  & $ {H}_{d}(1,2,-1/2)$  
 + $\bar{H}_{T}(\bar{3},1,+1/3)$ \\
 \hline 
 $~~~24_{A/H}$  & $(1,1,0)$ 
 + $(1,3,0)$ 
 + $(8,1,0)$ \\
 &+ $\chi(3,2,-5/6)$ 
 + $\bar{\chi}(\bar{3},2,5/6)$ \\
 \hline 
 $~~~ 50$  & $(3,1,-1/3)$ 
 + $ (\bar{3},2,-7/6)$ 
 + $ (1,1,-2)$\\
 &+ $ (6,1,4/3)$ 
 + $ (\bar{6},3,-1/3)$ 
 + $ (8,2,1/3)$ \\
 \hline 
 $~~~\overline{50}$  & $(\bar{3},1,+1/3)$ 
 + $ (3,2,+7/6)$ 
 + $ (1,1,+2)$  \\
& + $ (\bar{6},1,-4/3)$ 
 + $ (6,3,+1/3)$ 
 + $ (8,2,-1/2)$ \\
\end{tabular}
\end{ruledtabular}
\end{table}
\begin{table}[t!]
\caption{\label{qr} $U(1)_R$ and $U(1)_A$ charge assignments of superfields.
}
\begin{ruledtabular}
\begin{tabular}{ccc}
\textrm{ }&
\textrm{$q(U(1)_R)$}&
\textrm{$q(U(1)_A)~~~~ $}\\
\hline
$~~~~ W$ & $1$ & $0 ~~~~$ \\
\hline
$ ~~~~S$ & $1$ & $0 ~~~~$ \\
\hline
$ ~~~~10_i$ & $1/2$ & $q ~~~~$ \\
\hline
$ ~~~~\bar{5}_i$ & $1/2$ & $2q ~~~~$ \\
\hline
$~~~~ 1_i$ & $1/2$ & $0 ~~~~$ \\
\hline
$~~~~ 5_h$ & $0$ & $-2q~~~~ $\\
\hline
$~~~~ \bar{5}_h$ & $0$ & $-3q ~~~~$ \\
\hline
$ ~~~~50$ & $1$ & $3q~~~~ $ \\
\hline
$~~~~\overline{50}$ & $1$ & $2q~~~~ $ \\
\hline
$~~~~ X$ & $-1$ & $-5q~~~~ $ \\
\hline
$ ~~~~24_H$ & $0$ & $0~~~~ $\\
\end{tabular}
\end{ruledtabular}
\end{table}
In $R$-symmetric $SU(5)$ model, matter content of minimal supersymetric standard model (MSSM) and right handed neutrino (RHN) reside in $\bar{5}_i+10_i$ and $1_i$ dimensional representation respectively. Higgs and gauge superfields reside in $5_h, \bar{5}_h$ and $24_{A/H}$, along with a pair of $50$-plet, as shown in table \ref{sf}. $SU(5)$ breaks into gauge symmetry $SU(3)_c\times SU(2)_L\times U(1)_Y$ via vacuum expectation value (vev) in singlet direction of $24_H$ ($\langle 24_H\rangle \equiv M$). Pair of $50$-plets are added in the content  to incorporate missing doublet mechanism (MDM) in our R-symmetric $SU(5)$ model.

We have imposed an anomalous $U(1)_A$ symmetry through the introduction of a gauge singlet field $X$, whose non-zero vacuum expectation value (vev) is triggered by a D-term \cite{Dine:1987xk, Atick:1987gy, Dine:1987gj}, as expressed by the following equation:
\begin{eqnarray}
\langle X \rangle = \frac{g_A M_P}{8 \sqrt{3} \, \pi} ( Tr\ Q_A )^{1/2},
\end{eqnarray}
where the value of the reduced Planck mass $M_P = 2.4\times10^{18}$~GeV. In our numerical work, we consider $\langle X \rangle \sim 0.1 M_P$, a choice that is feasible for $g_A \sim 1$ and $Tr\ Q_A = -2q \sim 10$.  
Charges of suprfields under $U(1)_R$ and $U(1)_A$ symmetries are given in table \ref{qr}. These $U(1)_A$ charges for R-symmetric $SU(5)$ model are different from other $SU(5)$ models. It is important to note that $\mathbb{Z}_2$ matter parity is a subset of our $U(1)_R$ symmetry.

The superpotential under the above-mentioned symmetries becomes,
\begin{eqnarray}
W &=&
\kappa \ S\ (24_H^2-M^2) \nonumber \\
&+& y^{(u)} 10\ 10\ 5_h + y^{(l,d)} 10\ \bar{5}\ \bar{5}_h \nonumber \\
&+& \lambda \ \overline{50}\ \frac{24_H^2}{M_P}\ 5_h + \bar{\lambda}\ 50\ \frac{24_H^2}{M_P}\ \bar{5}_h+\lambda_{50}\ X\ 50\ \overline{50}\nonumber \\
&+& y^{(\nu)} \bar{5}\ 1\ 5_h+ \left(M_R+\eta \frac{24_H^2}{M_P}\right)\ 1\ 1.
\end{eqnarray}
The first line of superpotential is relevant for Higgs inflation where scalar standard model gauge singlet component $H$ of $24_H$ plays the role of inflaton. Second line in $W$ contains Yukawa interactions of charged superfields. In the third line of $W$, first two terms are nonrenormalizable level terms giving masses to color triplets in $5$ and $50$-plets, implementing the missing doublet mechanism. Third term in this line give heavy masses of the order of $\sim 10^{17}$ GeV to $50$-plet components, while keeping the model perturbative above GUT scale. Last line in $W$ explain neutrino masses via seesaw mechanism.

The $\mu 5_h\bar{5}_h$ term is forbidden by  $U(1)_A$ symmetry  upto all orders, leaving electroweak doublets in $5$-plets massless. Giudice-Masiero mechanism \cite{Giudice:1988yz} can generate MSSM $\mu$-term for light electroweak masses as a consequence of supergravity breaking in hidden sector. Superpotential of $SU(5)$ models gives the idea that Yukawa coupling $y^{(l,d)}$ can result in degenerate masses for down type quarks and charged lepton, this can be resolved by using nonrenormalizable terms. Generation dependant $U(1)_A$ charges \cite{Berezhiani:1996nu} and nonrenormalizable terms can explain fermion mass hierarchy in R-symmetric $SU(5)$ model.
\section{\label{sec:level2}Nonminimal Higgs inflation}

Superpotential terms involved in inflation are given by
\begin{equation}
    W\supset \kappa S(H^2-M^2),
\end{equation}
where $\kappa$ and $ M$ are real parameters. In our study we choose $M\sim10^{17}$ GeV to have a parameter space which prevents the decay of inflaton to Higgs color triplet as further discussed in section \ref{gcusec}. 
We consider a noncanonical K$\ddot{a}$hler potential $K=-3\ln\Phi$, where
\begin{equation}
\begin{split}
    \Phi& = 1-\frac{1}{3}(|S|^2+|H|^2)+\frac{\chi}{4}(H^2+h.c.) \\&
    +\frac{\sqrt{2}\gamma_3}{3}(S^2S^*+h.c.)+\frac{\gamma_4}{3}|S|^4.
    \end{split}
\end{equation}
Unless specified otherwise, the reduced Planck mass, $M_P$, is set to unity.

Note that the term involving $\gamma_3$ breaks R-symmetry at non-renormalizable level. This justification stems from the fact that quantum gravitational effects are expected to induce the breaking of all global symmetries \cite{HAWKING1987337,Lavrelashvili:1987jg,GIDDINGS1988854,COLEMAN1988643,GILBERT1989159,Banks_2011} $-$ a phenomenon we assumed in our model featuring a global $U(1)_R$ symmetry. These breaking effects are expected to appear at the nonrenormalizable level, characterized by a reduced Planck-mass suppression. The precise magnitude of these effects can be gauged by consistently embedding our model within a quantum theory of gravity, a task that falls beyond the scope of our present study.

The K\"ahler potential introduces leading-order R-symmetry breaking terms at the cubic level. The additional relevant R-symmetry breaking terms up to the quartic level, represented by $\alpha_i$ couplings, that contribute to $\Phi$ are expressed as follows:
\begin{equation}
\begin{split}
    \Delta \Phi& \supset \alpha_1(SH^2+h.c.)+\alpha_2(S^3+h.c.) \\
&+\alpha_3(SH^3+h.c.)+\alpha_4(S^2H^2+h.c.)\\&+\alpha_5(S^4+h.c.).
    \end{split}
\end{equation}
In a multifield treatment discussed later, both the terms with $\gamma_3$ and $\gamma_4$ couplings play crucial roles in shaping the potential for the correct inflationary track and generating primordial black holes. Numerical investigations indicate that R-symmetry breaking terms with an odd power of $S$ effectively emulate the role of the $\gamma_3$ coupling, while terms with an even power of $S$ support the role of the term with the $\gamma_4$ coupling. For simplicity, it is plausible to assume $\alpha_1 = \alpha_2 = \alpha_3 \sim \gamma_3$ and $\alpha_4 = \alpha_5 \sim \gamma_4$. Given that $S \ll 1$ near the pivot scale and $H<1$ during inflation, higher-order terms are expected to be suppressed. This suppression mitigates any apparent arbitrariness in the K\"ahler potential.

To scrutinize the potential existence of any undesirable $R$-symmetry-breaking terms in the superpotential, we exhaustively list all possible R-symmetry-breaking terms at the leading order of nonrenormalizable operators. These terms not only maintain invariance under $SU(5)$ gauge symmetry but are also consistent with $U(1)_A$ symmetry, as outlined below:
\begin{eqnarray}
&&S^4,\,\, S^2\ 24^2, \,\, S^3\ 1 , \,\, 24^4, \,\, 1\ 1\ 1\ 1, \,\, 10\ 10\ 5_h\ 1, \nonumber\\ 
&&10\ \bar{5}\ \bar{5}_h\ 1, \,\,  \bar{5}\ 1\ 5_h\ 1, \,\, 50\ X\ \overline{50}\ 1.
\end{eqnarray}
It is noteworthy that the presence of $U(1)_A$ symmetry significantly constrains the set of permissible terms. Given that the $S$ field remains close to the origin during inflation, terms such as $S^4$, $S^2\ 24^2$, and $S^3\ 1$ are adequately suppressed. The significance of the term $24^4$ lies in its role of conferring heavy masses to the octet and triplet components of the 24-plet, as elaborated in section V. Importantly, all other fields listed above involve matter fields that are assumed to be stabilized at zero during inflation, thereby posing no risk to our model. Consequently, the R symmetry-breaking terms do not present any threat to the viability of our model.

The singlet direction $s$ and the flat direction $h$ can be parameterized, along the scalar components of the multiplets, as:
\begin{equation}
      S=\frac{1}{\sqrt{2}}s, \qquad H=\frac{1}{\sqrt{2}}h.
\end{equation}
The scalar-gravity portion of the Lagrangian density is expressed in the following manner
\begin{equation}\label{e5}
      \mathcal{L}_J=\sqrt{-g_J}\left[\frac{\Phi\mathcal{R}_J}{2}-\frac{1}{2}g^{\mu\nu}_J\partial_{\mu}h\partial_{\nu}h-\frac{\zeta}{2}g^{\mu\nu}_J\partial_{\mu}s\partial_{\nu}s-V_J\right],
\end{equation}
where $g^{\mu\nu}_J$  is the inverse of the Jordan frame spacetime metric $g^J_{\mu\nu}$, $\mathcal{R}_J$ is the Ricci scalar in the Jordan frame and
\begin{align}
    \Phi&=1-\frac{s^2}{6}+\frac{\gamma_3}{3}s^3+\frac{\gamma_4}{12}s^4+\xi h^2,\\
    \zeta&=1-4\gamma_3s-2\gamma_4s^2,
\end{align}
where 
\begin{equation}
    \xi\equiv\frac{\chi}{4}-\frac{1}{6}.
\end{equation}
$V_J$ represents the F-term scalar potential in the Jordan frame.

In the Einstein frame Lagrangian becomes,
\begin{equation}
    \mathcal{L}_E=\sqrt{-g}\left[\frac{1}{2}\mathcal{R}-\frac{1}{2}G_{IJ}g^{\mu\nu}\partial_{\mu}\phi^I\partial_{\nu}\phi^J-V_E\right],
\end{equation}
where $\phi^I=(s,h)$ and $I=1,2$. The scalar potential is $V=\Phi^{-2}V_J.$ When considering the Einstein frame, the kinetic term for the scalar fields incorporates a field-space metric that is not trivial\footnote{In~\cite{Giare:2023kiv} the authors introduced a numerical method specifically designed for investigating generic
multifield models of inflation with a non-trivial field space metric.}\cite{Kaiser_2013},
\begin{equation}
    G_{IJ}=\frac{1}{2f}\left[\mathcal{G}_{IJ}+\frac{3}{f}f_{,I}f_{,J}\right],
\end{equation}
where $\mathcal{G}_{IJ}$ is the field-space metric in Jordan frame and $f_{,I}\equiv\partial f/\partial\phi^I$ with $f=1/2 \, e^{-K/3}$.
The Christoffel symbol for the field space is computed from the metric $G_{IJ}$ as,
\begin{equation}
    \Gamma^I_{JK}=\frac{1}{2}G^{IL}\left(G_{LK,J}+G_{LJ,K}-G_{JK,L}\right),
\end{equation}
where $G^{IJ}$ is the invesre of $G_{IJ}$.

The F-term supergravity (SUGRA) scalar potential in Einstein frame is defined as,
\begin{equation}
    V_F=e^{K/M_P^2}\left(K_{ij}^{-1}D_{z_i}WD_{z^*_j}W^*-3M_P^{-2}\left|W\right|^2\right),
\end{equation}
where,
\begin{equation}
    D_{z_i}W\equiv\frac{\partial W}{\partial z_i}+M_P^{-2}\frac{\partial K}{\partial z_i},\quad K_{ij}\equiv\frac{\partial^2K}{\partial z_i\partial z^*_j},
\end{equation}
$D_{z^*_i}=(D_{z_i}W)^*, z_i\in \{H,S\}$, and $z^*_i$ is the conjugate. 

\begin{figure*}[t!]\centering
{\includegraphics[width=0.45\textwidth]{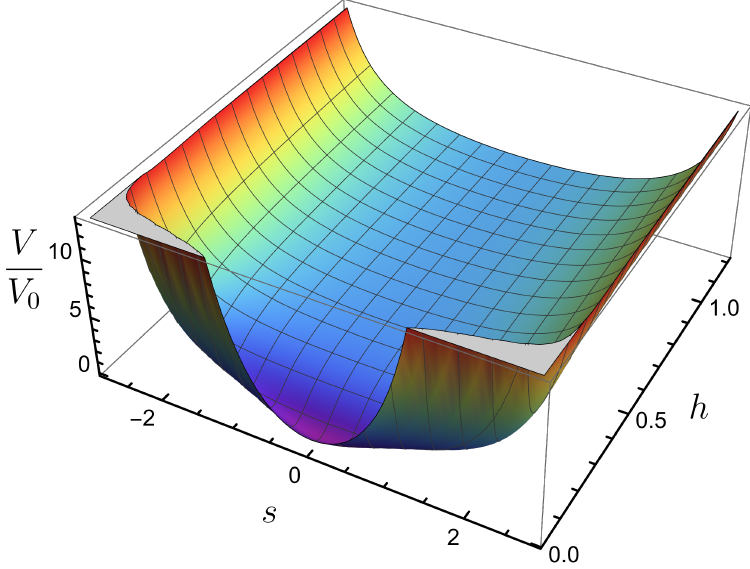}}\qquad
{\includegraphics[width=0.45\textwidth]{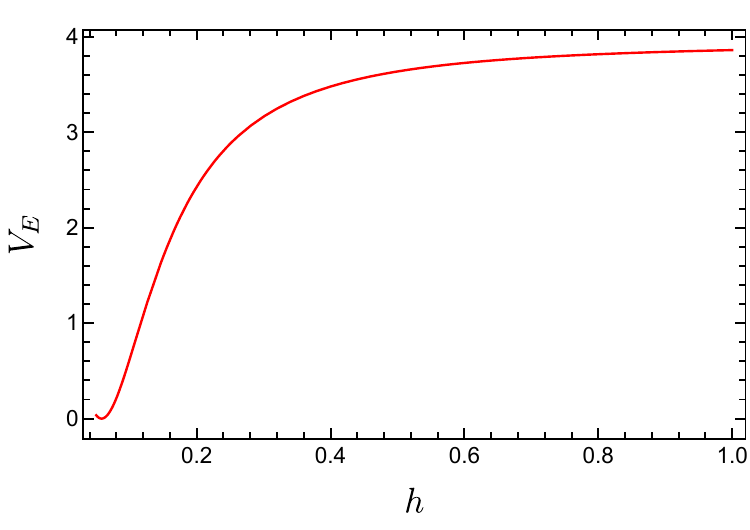}}
\caption{\label{fig:3D}The normalized scalar potential in the Einstein frame (left) and its $s=0$ slice for $M=10^{17}, \xi=150$ and $\gamma_4=0.045$ (right).}
\end{figure*}

In order to examine how background-order quantities and fluctuations evolve in these models, we utilize the gauge-invariant multifield approach of Refs. \cite{Kaiser_2013,Peterson_2011,Gordon_2000,Nakamura_1996,Gong_2011}. We consider perturbations around a spatially flat Friedmann-Lemaitre-Robertson-Walker spacetime. Each scalar field may be decomposed into its vacuum expectation value and a fluctuation that varies across space,
\begin{equation}
    \phi^I(x^{\mu})=\varphi^I(t)+\delta\phi^I(x^{\mu}).
\end{equation}
The magnitude of the velocity of the background fields can be written as,
\begin{equation}
    |\Dot{\varphi}|\equiv\Dot{\sigma}=\sqrt{G_{IJ}\Dot{\varphi}^I\Dot{\varphi}^J},
\end{equation}
the overdot is the derivative with respect to time, hence unit vector becomes,
\begin{equation}
    \hat{\sigma}^I\equiv\frac{\Dot{\varphi}^I}{\Dot{\sigma}},
\end{equation}
which points along the direction of background fields of motion in field space. The notation,
\begin{equation}
    \hat{s}^{IJ}\equiv G^{IJ}-\hat{\sigma}^I\hat{\sigma}^I,
\end{equation}
projects onto the subspace of the field-space manifold perpendicular to the background fields' motion. The covariant turn-rate vector is defined as,
\begin{equation}
    \omega^I\equiv\mathcal{D}_t\hat{\sigma}^I,
\end{equation}
where 
$\mathcal{D}_tA^I\equiv\Dot{\varphi}^J\triangledown_JA^I$, for any field-space vector $A^I$ and  $\triangledown_J=\partial_JA^I+\Gamma^I_{JK}A^K$, is the covariant derivative. The slow-roll parameters are defined as,
\begin{eqnarray}
    \epsilon&\equiv&-\frac{\Dot{H}}{H^2},\\
    \eta&\equiv& 2\epsilon-\frac{\Dot{\epsilon}}{2H\epsilon}.
\end{eqnarray}

The background evolution is described by the equations of motion for multifields as,
\begin{equation}
    \mathcal{D}_t\dot{\varphi}^I+3H\dot{\varphi}^I+G^{IJ}\triangledown_{J}V(\varphi^K)=0,
\end{equation}
together with the Friedmann equation,
\begin{equation}
    3H^2=\frac{1}{2}G_{IJ}\dot{\varphi}^I\dot{\varphi}^J+V(\varphi^K).
\end{equation}
Dynamics of the perturbations is found by solving the equations,
\begin{align}\label{e20}
    \ddot{Q}_{\sigma}+3H\dot{Q}_{\sigma}&+\left[\frac{k^2}{a^2}+\mathcal{M}_{\sigma\sigma}-\omega^2-\frac{1}{a^3}\frac{d}{dt}\left(\frac{a^3\dot{\sigma}^2}{H}\right)\right]Q_{\sigma}\notag\\
    &=2\frac{d}{dt}\left(\omega Q_{s}\right)-2\left(\frac{V_{,\sigma}}{\dot{\sigma}}+\frac{\dot{H}}{H}\right)\left(\omega Q_{s}\right)\\
     \Ddot{Q}_s+3H\Dot{Q}_s&+\left[\frac{k^2}{a^2}+\mathcal{M}_{ss}+3\omega^2\right]Q_s\notag\\
    &=4\frac{\omega}{\Dot{\sigma}}\frac{\Dot{H}}{H}\left[\frac{d}{dt}\left(\frac{H}{\Dot{\sigma}}Q_{\sigma}\right)-\frac{2H}{\Dot{\sigma}}\omega Q_{s}\right],
\end{align}
where $Q_{\sigma}$ and $Q_{s}$ are respectively the adiabatic and entropic perturbations, and,
\begin{align}
    \mathcal{M}_{\sigma\sigma}&\equiv\hat{\sigma}_I\hat{\sigma}^J\mathcal{M}^I_J,\\
    \mathcal{M}_{ss}&\equiv\hat{s}_I\hat{s}^J\mathcal{M}^I_J,\\
    \mathcal{M}^I_J&\equiv G^{IK}\triangledown_J\triangledown_KV-\mathcal{R}^I_{LMJ}\Dot{\varphi}^L\Dot{\varphi}^M,\\
    V_{,\sigma}&\equiv \hat{\sigma}^I\triangledown_IV.
\end{align}
Here, $\mathcal{R}^I_{LMJ}$ is the field space Riemann tensor. The curvature and isocurvature perturbations are defined as follows,
\begin{equation}
    \mathcal{R}=\frac{H}{\Dot{\sigma}}Q_{\sigma}, \qquad \mathcal{S}=\frac{H}{\Dot{\sigma}}Q_s.
\end{equation}
The effective mass-squared of the entropy perturbations is defined as,
\begin{equation}\label{mu}
    \mu_s^2\equiv \mathcal{M}_{ss}+3\omega^2.
\end{equation}

\subsection{Effective single field inflation}
\begin{figure*}[t!]\centering
\subfloat[\label{xi}]{\includegraphics[width=0.53\textwidth]{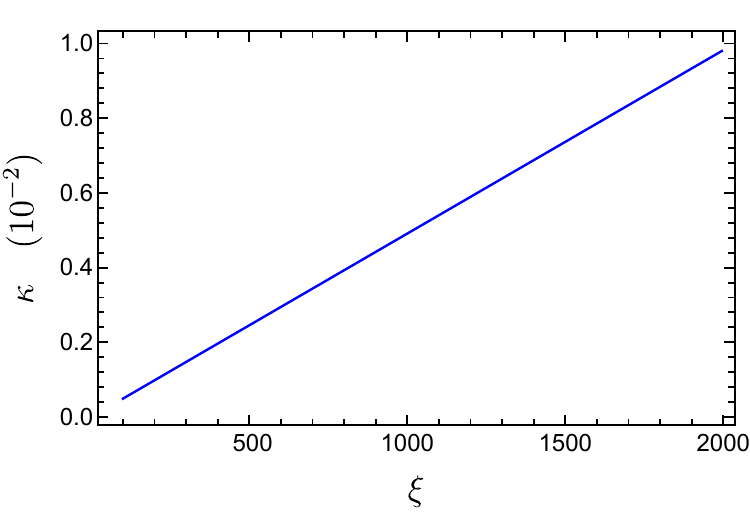}}\qquad
\subfloat[\label{si}]{\includegraphics[width=0.37\textwidth]{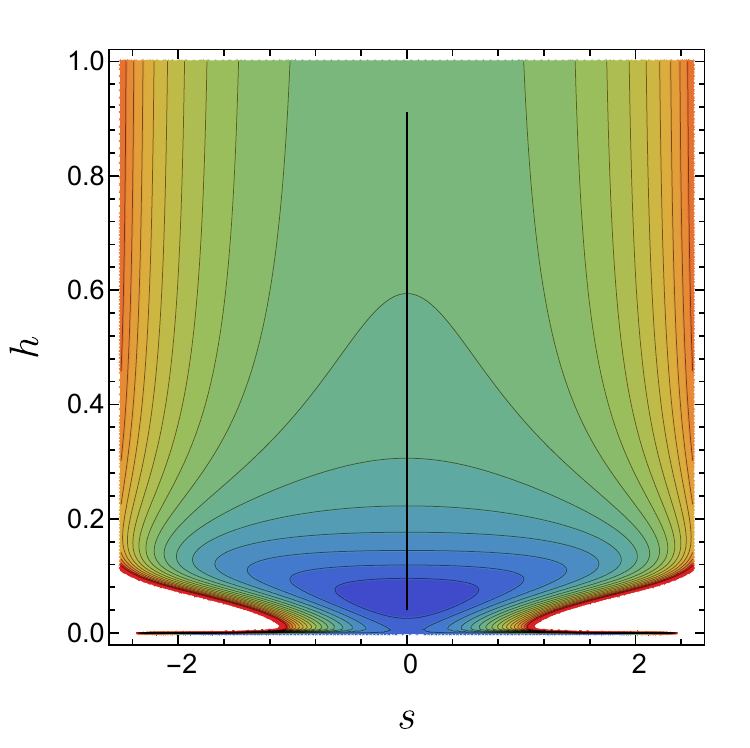}}
\caption{\label{xisi}(a) Relation between $\kappa$ and $\xi$. (b)  The inflaton trajectory, since $s$ is stabilized at zero the inflaton follows a straight trajectory.}
\end{figure*}

The $s$ field is stabilized at $s=0$ for $\gamma_3=0$ and large values of $\gamma_4$, the scenario then reduces to effective single field inflation. The 3D potential in the Einstein frame along with the $s=0$ slice is shown in Fig.~\ref{fig:3D}. In this case we have five free parameters $\gamma_4, \xi, \kappa$, initial field value ($h_\text{ini},s_\text{ini}$) and the initial field velocities ($\dot{h}_\text{ini},\dot{s}_\text{ini}$). The initial velocities are determined by the slow roll background equations of motion whcih is a good approximation for large $h$. The initial field value is chosen as $0.95$, enough to achieve 60 efolds. The $\kappa$ parameter appears as an overall factor in the potential and is chosen to match the curvature power spectrum at the pivot scale. Now we are left with two parameters, $\xi$ and $\gamma_4$. The relation between $\xi$ and $\kappa$ is depicted in Fig.~\ref{xi}. We choose $\xi=150$ and $\gamma_4=0.045$. In Fig.~{\ref{si}} the inflaton trajectory for the above parameters is presented and it can be seen that the inflaton follows a straight trajectory. The predicted values for the inflationary observables are:
\begin{equation}
    n_s\simeq0.965,\qquad r\simeq10^{-3}, \qquad -\frac{dn_s}{d\ln k}\simeq10^{-4}.
\end{equation}
These predictions of the model for inflationary observables are in good agreement with Planck 2018 data~\cite{planck2020}. This type of scenario has been considered recently in \cite{Abid_2021} probing flipped $SU(5)$, in \cite{ahmed2023probing} for $\chi SU(5)$, in \cite{Arai_2011,Kawai_2016} for $SU(5)$ without R-symmetry, and in \cite{Masoud_2019} with additional $U(1)_R \times Z_n$ symmetry. The monopoles produced due to $SU(5)$ symmetry breaking are inflated away during inflation. The shifted \cite{Khalil_2011} and new \cite{Rehman_2020} inflation models of R-symmetric $SU(5)$ have also been employed to avoid the same monopole problem.
Another interesting possibility appears in \cite{Tavartkiladze:2019svd} where inflaton emerges as a superposition of the Higgs, squark and slepton.

\subsection{Reheating and Leptogenesis}
The inﬂaton starts oscillating about the minimum after inflation ends. During this process the inﬂaton can decay into the lightest right handed neutrino, $N$, via $W \supset 24_H^2 1_i 1_j$ with the partial decay width given by \cite{Pallis_2011},
\begin{equation}
    \Gamma_{N}=\frac{1}{64\pi}\left[ \frac{M_{N}}{M}\frac{\Omega_{0}^{3/2}}{J_0}\left(1-\frac{12\xi M^2}{M_P^2}\right)\right]^2 m_\text{inf}\sqrt{1-\frac{4M_N^2}{m_\text{inf}^2}},
\end{equation}
where $m_\text{inf}=\sqrt{2}\kappa M/\Omega_0 J_0$ is the inﬂaton mass with,
\begin{equation}
\begin{split}
  J_0&=\sqrt{\frac{1}{\Omega_0}+\frac{6\xi^2M^2}{M_P^2\Omega_0^2}}, \\
    \Omega_0&=1+\frac{4\xi M^2}{M_P^2},
\end{split}    
\end{equation}
and $M_{N}$ is the mass of the RHN. The predicted range of the
RHN mass is $M_{N}\simeq 10^{11}-10^{12}$ GeV, with an inﬂaton mass, $m_\text{inf}\sim 10^{13}$ GeV.

The decay of inflaton to color triplet pairs via $W\supset \lambda \overline{50}\, 24_H^2\,5_h/M_P + \bar{\lambda} {50}\,24_H^2\, \bar{5}_h/M_P$ has been suppressed by selecting a parametric range which renders the triplet heavier than inflaton. Otherwise, the reheating temperature could be very high, leading to the problem of gravitino overproduction problem \cite{Ellis:1984eq, Khlopov:1984pf}. 

The inﬂaton can also decay via the top Yukawa coupling, $y_{33}^{\mu,\nu}Q_3L_3H_u$, in a supergravity framework as described in \cite{Endo_2006,Endo_2007}. In no-scale like SUGRA models, this decay width is given by the following expression \cite{Pallis_2011},
\begin{equation}
    \Gamma_y=\frac{3}{128\pi^3}\left(\frac{6\xi y \Omega^{3/2}_0}{J_0}\right)^2\left(\frac{M}{M_P}\right)^2\left(\frac{m_\text{inf}}{M_P}\right)^2,
\end{equation}
where $y=y_{33}$ is the top Yukawa coupling. The reheat temperature $T_r$ is related to the total decay width of the inﬂaton $\Gamma$ namely,
\begin{equation}
    T_r=\left(\frac{72}{5\pi^2g_{*}}\right)^{1/4}\sqrt{\Gamma M_P},\quad \Gamma= \Gamma_N+ \Gamma_y,
\end{equation}
where $g_{*}\simeq228.75$ for MSSM. Under the assumption of a conventional thermal history, we can express the number of e-folds $N_0$ in relation to the reheat temperature $T_r$ as \cite{Liddle_2003},
\begin{equation}
    N_0=53+\frac{1}{3}\ln{\left[\frac{T_r}{10^9\text{GeV}}\right]}+\frac{2}{3}\ln{\left[\frac{\sqrt{\kappa}M}{10^{15}\text{GeV}}\right]}.
\end{equation}

In the context of supergravity-based inflationary models, one has to deal with the gravitino overproduction problem \cite{khlopov1984easy,ellis1984cosmological}. 
This problem places stringent constraints on the reheat temperature $T_r$ in relation to the gravitino mass $m_{3/2}$. 
In scenarios involving gravity-mediated supersymmetry breaking for unstable gravitinos with a mass exceeding 10~TeV, the reheat temperature becomes nearly independent of the gravitino mass. On the other hand, for stable gravitinos, the reheat temperature is constrained to be $T_r\lesssim 10$~GeV, as discussed in \cite{Kawasaki_2005,Kawasaki_2008,Kawasaki_2018}.
Consequently, the predicted range for the reheat temperature, $T_r \sim 4.3\times10^{8}-4.6\times 10^{9}$~GeV is consistent with the constraints imposed by the gravitino problem.

The generation of lepton asymmetry through inflaton decay involves a partial conversion into baryon asymmetry via sphaleron processes \cite{KHLEBNIKOV1988885,PhysRevD.42.3344}. To account for the observed baryon asymmetry, we explore the non-thermal leptogenesis scenario \cite{Senoguz_2004}. In this context, the ratio of lepton number to entropy density, denoted as $n_L/s$, can be expressed as,
\begin{equation}
    n_L/s\simeq\frac{3T_r}{2m_{inf}}\left(\frac{\Gamma_N}{\Gamma}\right)\epsilon_1.
\end{equation}
Here, $\epsilon_1$ represents the CP asymmetry factor, generated during the out-of-equilibrium decay of the lightest right-handed neutrino, $N$. Assuming a normal hierarchical pattern for observed neutrinos results in \cite{hamaguchi2002cosmological},
\begin{figure*}[t!]\centering
\subfloat[\label{t1}]{\includegraphics[width=0.45\textwidth]{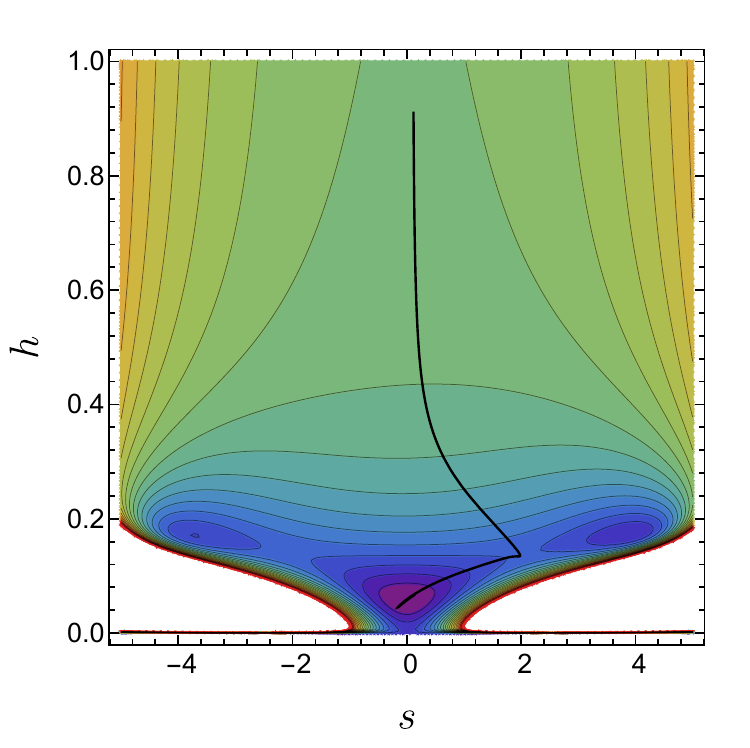}}\qquad
\subfloat[\label{t2}]{\includegraphics[width=0.45\textwidth]{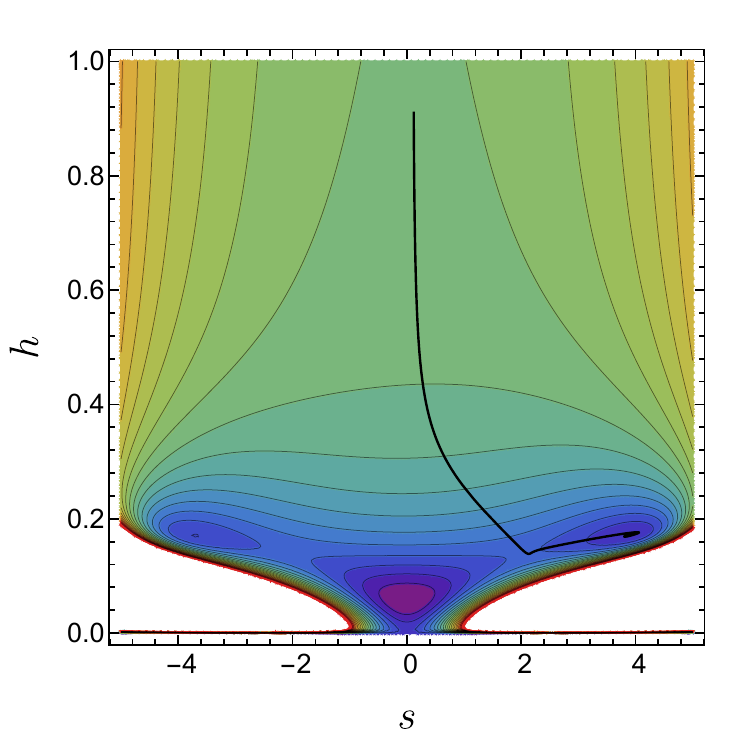}}
\caption{\label{inflaton_t}Examples of inflationary trajectories: (a) for ($\gamma_4=0.007,\gamma_3=0.00066$) and (b) for ($\gamma_4=0.007,\gamma_3=0.00068$).}
\end{figure*}
\begin{equation}
    \epsilon_1\simeq-\frac{3m_{N_3} M_{N}}{8\pi\nu_u^2}\delta_{eff},
\end{equation}
where $\delta_{eff}$ represents the effective CP-violating phase, $\nu_u$ denotes the vacuum expectation value of the up-type electroweak Higgs doublet, and $m_{N_3}$ corresponds to the mass of the heaviest left-handed light neutrino.
With the above expression of $\epsilon_1$, the ratio of lepton number to entropy density becomes,
\begin{equation}
\begin{split}
    n_L/s&\simeq3.9\times 10^{-10}\left(\frac{\Gamma_{N}}{\Gamma}\right)\\ &\times\left(\frac{T_{r}}{m_{inf}}\right)\left(\frac{M_{N}}{10^6 \hspace{2pt}\text{GeV}}\right)
    \left(\frac{m_{N_3}}{0.05\hspace{2pt}\text{eV}}\right)\delta_{eff}.
\end{split}
\end{equation}
The effective CP-violating phase $|\delta_{\text{eff}}|$ is constrained to be less than or equal to one. We consider the mass of the heaviest light neutrino to be $m_{N_3}=0.05$~eV. In order to generate the required lepton asymmetry, $n_L/s\approx 6.12\times10^{-10}$, we make the assumption of a hierarchical neutrino mass pattern $M_N\ll M_{N_2}\ll M_{N_3}$ with the condition that $M_N$ exceeds the reheat temperature, i.e., $M_N > T_r$.

\section{\label{PBHGW}PBHs and gravitational waves}

A positive quartic term ($\gamma_4 > 0$) is required to stabilize $s$ during inﬂation. As $\gamma_4$ decreases, additional minima at $s \neq 0$ appear in the potential. Introducing the parameter $\gamma_3$ tilts the potential asymmetrically in the $s$-field direction. By choosing an appropriate $\gamma_3$ value, the inflaton trajectory can pass through the saddle point between the true vacuum at $s=0$ and the false vacuum at $s \neq 0$. In a recent paper \cite{Kawai_2023}, a superpotential of the form, $y S X \bar{X}$, has been studied in a model-independent way with a specific choice of parameters involved. It is important to mention here that for the realization of the present model, a completely different choice of parameters is needed as described below.

In Fig. \ref{inflaton_t}, we present two examples illustrating the trajectory of the inflaton for different values of $\gamma_3$. For these examples, we fix $\gamma_4=0.007$ and $\xi=150$. $\gamma_3=0$, is the same effective single field case and the inflaton follows a straight trajectory. However, for a significantly negative value of $\gamma_3$, the inflaton trajectory leads to a false vacuum state characterized by $s\neq0$.

To achieve a desirable scenario, it is important to choose an appropriate $\gamma_3$ value. In Fig.~\ref{t1}, we observe that the inflaton trajectory successfully crosses the saddle point between the false vacuum and the true vacuum located at $s=0$. This crossing is crucial as it triggers a suppression of the first Hubble slow roll parameter, $\epsilon$. This behavior indicates the onset of an ultra-slow roll regime. The presence of the ultra-slow roll regime is significant as it suggests a potential enhancement of curvature perturbations. Consequently, it opens up the possibility for the formation of PBHs. By carefully selecting the appropriate values for the parameters, we can explore the dynamics of the inflaton trajectory and its implications for the formation of PBHs.

\begin{figure*}[t!]\centering
\subfloat[\label{fig:mu}]{\includegraphics[width=0.45\textwidth]{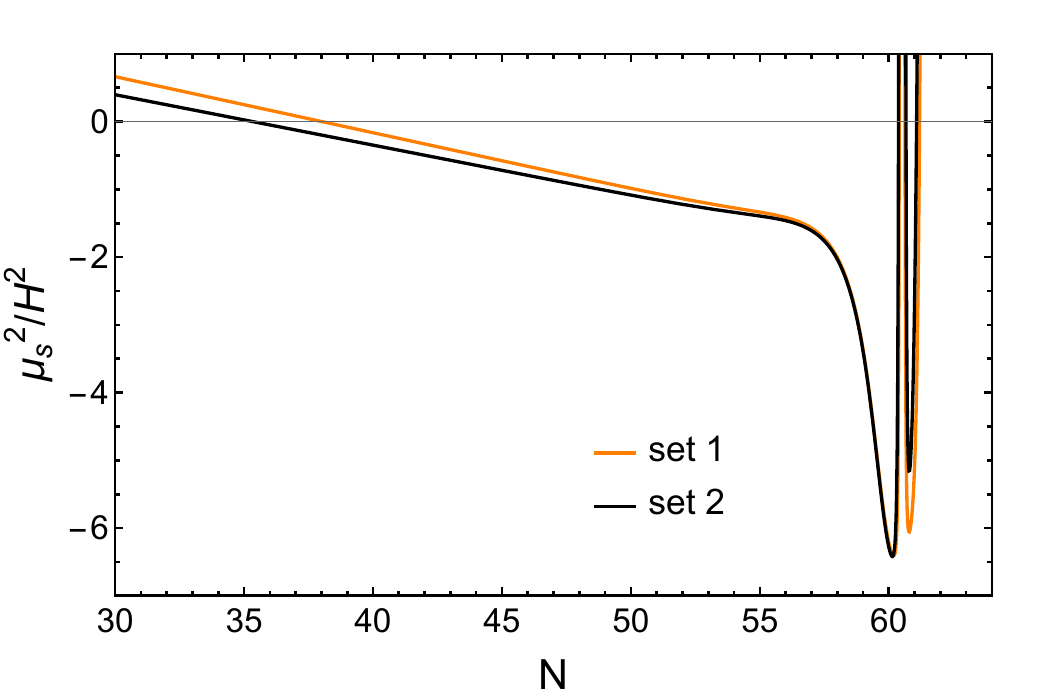}}\quad
\subfloat[\label{fig:fields}]{\includegraphics[width=0.45\textwidth]{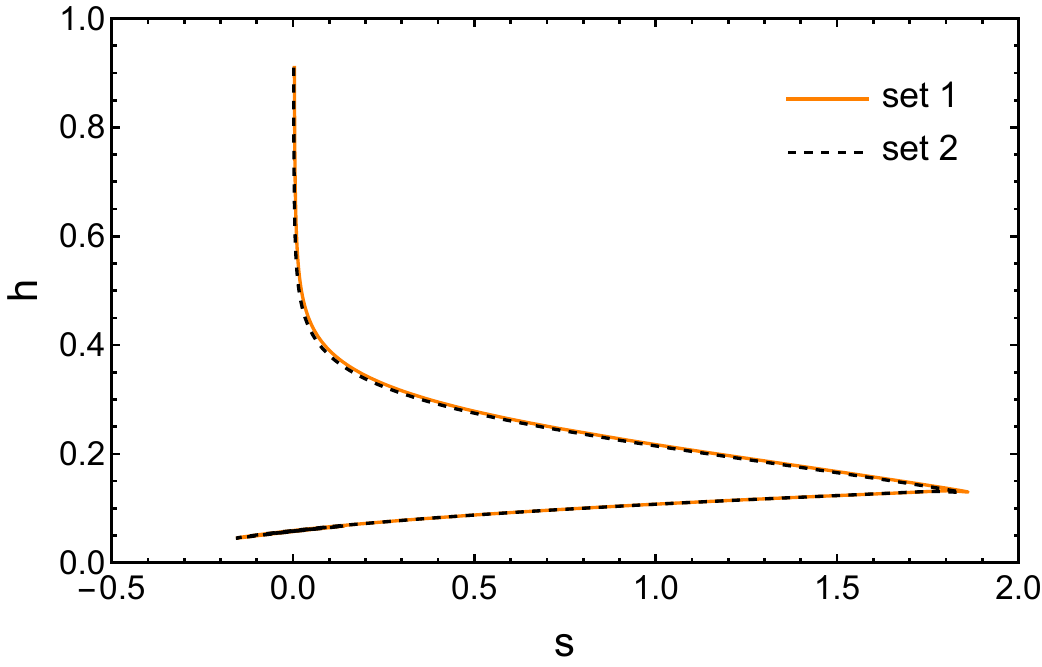}}
\caption{\label{fig:mufields} (a) The effective entropic mass squared becomes tachyonic ($\mu^2<0$) for sufficient number of efolds. (b) Inflationary trajectories turing in the field space. These trajectories results a non-zero turn-rate $\omega\neq0$.}
\end{figure*}

Next, the equation for the adiabatic modes (\ref{e20}) shows that these modes are sourced by the product of the turn rate $\omega$ and the isocurvature modes $Q_s$. If the turn-rate is non-zero $\omega\neq0$ and the amplitude of the isocurvature modes $Q_s$  is sufficiently large, then the isocurvature modes can source the adiabatic modes. Therefore, for realizing the multifield amplification mechanism several factors have to work in a synchronized way. First, the growth of isocurvature modes require a sufficiently long inflationary phase with a tachyonic effective isocurvature mass. Next, transmitting these enhancements to the adiabatic modes necessitates that the inflaton follows a curved trajectory within the field space, resulting in a non-zero turn rate. To prevent excessive amplification, the sourcing process must come to an end, either through the turn rate diminishing or by suppressing the isocurvature modes \cite{PhysRevD.103.083518}.

In our case the potential satisfies all these requirements. The effective mass squared for the isocurvature modes becomes negative for sufficient long period as shown in the Fig.~\ref{fig:mu} for the sets of parameters given in table~\ref{tab:table1}. The trajectories of the fields are turning (Fig.~\ref{fig:fields}) in the field space resulting a non-zero turn-rate. 

We provide the numerical results for the sets of parameters in table~\ref{tab:table1}. The predictions for inflationary observables are also given in the same Table. We compute the curvature power spectrum $P_{\zeta}(k)$, using a publicly available package of transport method \cite{dias2016computing}, to quantify the enhancement in the power spectrum.
In Fig.~\ref{fig:ps}, we present the curvature power spectrum as a function of the wavenumber for the sets of parameters in table~\ref{tab:table1}.

\begin{table}[t!]
\caption{\label{tab:table1}%
Sets of parameters.
}
\begin{ruledtabular}
\begin{tabular}{cccccc}
\textrm{}&
\textrm{$\xi$}&
\textrm{$\gamma_4$}&
\textrm{$\gamma_3$}&
\textrm{$n_s$}&
\multicolumn{1}{c}{\textrm{$r$}}\\
\colrule
set 1 & $150$ & $0.0034$ & $-7.94\times 10^{-6}$ & 0.962 & 0.0041\\
\hline
set 2 & $150$ & $0.00306$ & $-3.246\times 10^{-6}$ & 0.966 & 0.0040\\
\end{tabular}
\end{ruledtabular}
\end{table}

\subsection{Primordial Black Holes}
The enhanced curvature perturbations can lead to the formation of PBHs through gravitational collapse during horizon reentry. In this section we calculate the mass of PBHs and their fractional energy density abundances. We will assume that the PBHs are formed during the radiation dominated epoch.

The mass of a PBH is determined by the mass of the horizon at the time when the perturbation corresponding to the PBH enters the horizon. The relationship between the scale of the perturbation and the mass of the PBH, at the time of formation, is expressed as follows,
\begin{equation}
M_\text{PBH}=\gamma M_{H,0}\Omega_{rad,0}^{1/2}\left(\frac{g_{*,0}}{g_{*,f}}\right)^{1/6}\left(\frac{k_0}{k_f}\right)^2,
\end{equation}
where $M_{H,0}=\frac{4\pi}{H}$ represents the mass of horizon, $\Omega_{rad}$ is used to denote the energy density parameter for radiation, while $g_*$ represents the effective degrees of freedom. Furthermore, the $f$ and $0$ in the subscript corresponds to the time of formation and today respectively. $\gamma$ represents the ratio between the PBH mass and the horizon mass, which is estimated as $\gamma\simeq3^{-3/2}$ in the simple analytical result \cite{carr1975primordial}.

The energy density of the PBHs today can be obtained by redshifting that at the formation time, namely $\rho_\text{PBH,0}=\rho_\text{PBH,f}(a_\text{f}/a_0)^3\approx \gamma \beta \rho_\text{rad,f}(a_\text{f}/a_0)^3$, since the PBHs behave as matter. Here $\beta$ denotes the mass
fraction of Universe collapsing in PBH mass.

The mass fraction $\beta$ is evaluated using the Press-Schechter method assuming that the overdensity $\delta$ follows a gaussian probability distribution function. The collapse of PBH is determined by a threshold value denoted as $\delta_c$. Thus the mass fraction is given via the integral,
\begin{equation}
    \beta=\int_{\delta_c}^{\infty}d\delta\frac{1}{\sqrt{2\pi\sigma^2}}\exp\left(-\frac{\delta^2}{2\sigma^2}\right),
\end{equation}
where $\sigma$ is the variance of curvature perturbation that is related to the co-moving wavenumber \cite{Young_2014}.
\begin{equation}
    \sigma^2=\frac{16}{81}\int^{\infty}_{0}d\ln q (q k^{-1})^4W^2(q k^{-1}))\mathcal{P}_{\zeta}(q),
\end{equation}
where $W^2(q k^{-1})$ is a window function that we approximate with a Gaussian distribution function,
\begin{equation}
    W(q k^{-1})=\exp\left[-\frac{1}{2}(q k^{-1})^2\right].
\end{equation}
For $\delta_c$ we assume values in the range between $0.4$ and $0.6$ \cite{Harada_2013,Musco_2009,Musco_2005,Escriv__2020,Escriv__2021,Musco_2021}.

The total abundance is $\Omega_\text{PBH,tot}=\int d\ln M_\text{PBH} \Omega_\text{PBH}$, with $\Omega_\text{PBH}$ expressed in terms of $f_{\text{PBH}}$ which is given by,
\begin{align}
    f_{\text{PBH}}&\equiv\frac{\Omega_{\text{PBH,0}}}{\Omega_{\text{CDM,0}}}\nonumber\\
    &\approx\left(\frac{\beta(M)}{8.0\times10^{-15}}\right)\left(\frac{0.12}{\Omega_{\text{CDM,0}}h^2}\right)\left(\frac{\gamma}{0.2}\right)^{3/2}\nonumber\\&
    \times \left(\frac{106.75}{g_{*,f}}\right)^{1/4} \left(\frac{M_\text{PBH}}{10^{20} \text{g}}\right)^{-\frac{1}{2}},
\end{align}
where  $\Omega_{CDM,0}$ is the today's density parameter of the cold dark matter and $h$ is the rescaled Hubble rate today.

\begin{figure*}[t!]\centering
\subfloat[\label{fig:ps}]{\includegraphics[width=0.45\textwidth]{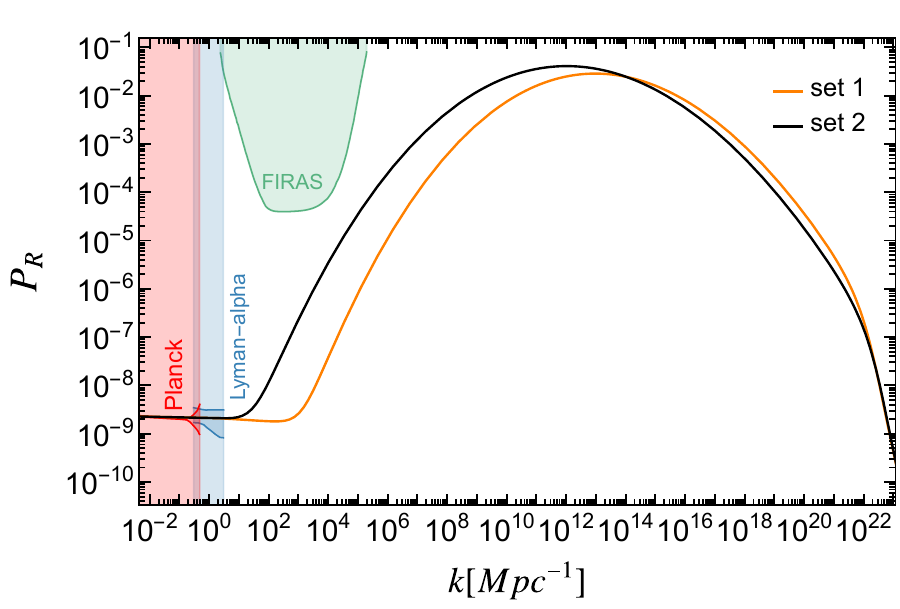}}\qquad
\subfloat[\label{fig:PBHDM}]{\includegraphics[width=0.45\textwidth]{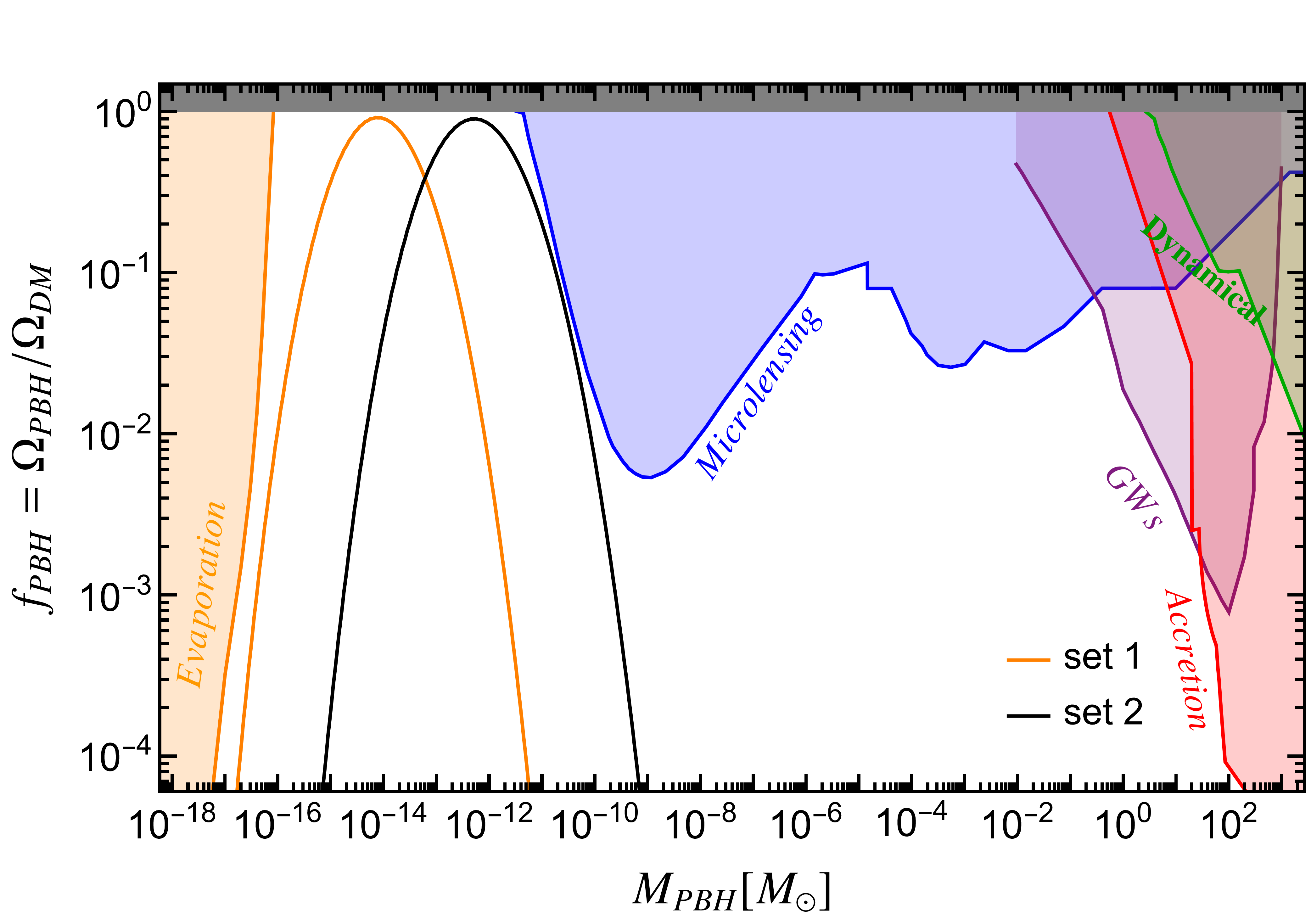}}\qquad
\caption{\label{fig:PSPBHDM} (a) The enhancement of the curvature power spectrum. (b)  The fractional abundance of PBHs as a function of their mass. Orange curve correspond to set 1 and the black to set 2, as given in Table \ref{tab:table1}. For the bounds we consulted \cite{Green_2021}.}
\end{figure*}

The fractional abundance of the primordial black holes as a function of their mass is shown in Fig.~\ref{fig:PBHDM} We have used the benchmark point parameters of table~\ref{tab:table1}. In both cases PBHs can explain dark matter in totality. The shaded regions correspond to the observational constraints \cite{Carr_2010,Carr_2021,Green_2021}.

\subsection{Scalar-induced gravitational wave}
The GWs energy density within the subhorizon regions can be expressed as \cite{Maggiore_2000},
\begin{equation}\label{4.12}
    \rho_{\text{gw}}=\frac{\langle \overline{\partial_lh_{ij}\partial^lh^{ij}}\rangle}{16a^2},
\end{equation}
here the average over oscillation is denoted by the overline and $h_{ij}$ is the tensor mode. We can decompose $h_{ij}$ as,  
\begin{equation}\label{4.13}
    h_{ij}(t,\textbf{x})=\int \frac{d^3\textbf{k}}{(2\pi)^{3/2}}\left( h_\textbf{k}^{+}(t)e_{ij}^{+}(\textbf{k})+h_{\textbf{k}}^{\times}(t)e_{ij}^{\times}(\textbf{k})\right)e^{i\textbf{k}.{x}},
\end{equation}
with $e_{ij}^{+}(\textbf{k})$ and $e_{ij}^{\times}(\textbf{k})$ being the polarization tensors and can be expressed as,  
\begin{align}
 e_{ij}^{+}(\textbf{k})&=\frac{1}{\sqrt{2}}\left(e_{i}(\textbf{k})e_{j}(\textbf{k})-\bar{e}_i(\textbf{k})\Bar{e}_j(\textbf{k})\right),\\
  e_{ij}^{\times}(\textbf{k})&=\frac{1}{\sqrt{2}}\left(e_{i}(\textbf{k})\Bar{e}_{j}(\textbf{k})-\bar{e}_i(\textbf{k})e_j(\textbf{k})\right),
\end{align}
here $e_{i}(\textbf{k})$ and $\Bar{e}_{j}(\textbf{k})$ are orthongal to eachother. Expression (\ref{4.13}) can be used in (\ref{4.12}), which yields, 
\begin{equation}
    \rho_{\text{gw}}(t)=\int d\ln k\frac{\overline{\mathcal{P}_h{(t,k)}}}{2}\left(\frac{k}{2a}\right)^2.
\end{equation}
In this expression $\mathcal{P}_{h}\equiv\mathcal{P}_h^{+,\times}$ and can be defined as, 
\begin{align}
    \langle h_{\textbf{k}}^{+}(t)h_{\textbf{q}}^{+}(t)\rangle&=\frac{2\pi^2}{k^3}\mathcal{P}_{h}^{+}(t,k)\delta^3(\textbf{k}+\textbf{q}),\\
     \langle h_{\textbf{k}}^{\times}(t)h_{\textbf{q}}^{\times}(t)\rangle&=\frac{2\pi^2}{k^3}\mathcal{P}_{h}^{\times}(t,k)\delta^3(\textbf{k}+\textbf{q}).
\end{align}

It should be noted that $\mathcal{P}_h^{+}(t,k)=\mathcal{P}_h^{\times}(t,k)$. The energy density parameter for gravitational wave is defined as follows,
\begin{equation}
    \Omega_{\text{gw}}(t,k)\equiv\frac{\rho_{\text{gw}}(t,k)}{\rho_{\text{crit}}}=\frac{1}{24}\left(\frac{k}{aH}\right)^2\overline{\mathcal{P}_h(t,k)},
\end{equation}
where we have omitted the polarization index. 

\begin{figure}[t!]\centering
\includegraphics[width=0.48\textwidth]{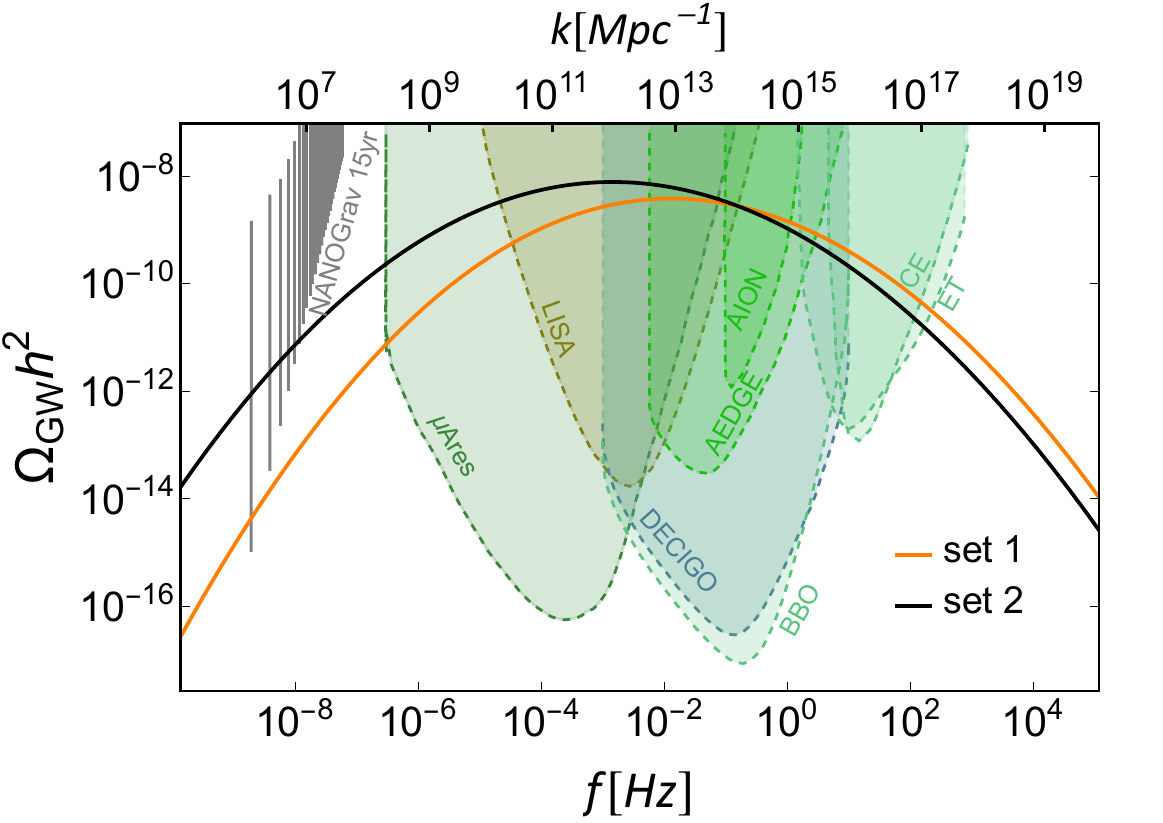}
\caption{\label{fig:gw} The gravitational waves spectrum together with the sensitivity curves of the various existing and forthcoming GW experiments.}
\end{figure}

The following tensor perturbation equation should be solved for obtaining the tensor power spectrum  \cite{Baumann_2007,Kohri_2018,Inomata_2017},
\begin{equation}
    h^{\prime\prime}_{\textbf{k}}+2aHh_{\textbf{k}}^{\prime}+k^2h_{\textbf{k}}=4S_{\textbf{k},}
\end{equation}
where derivatives with respect to conformal time are denoted by the prime. The source term $S_k$  is expressed as,
\begin{equation}
    \begin{split}
        S_k=\int&\frac{d^3q}{\sqrt[3]{(2\pi)}}q_iq_je_{ij}(\textbf{k})\left.\biggr[2\Phi_{\textbf{q}}\Phi_{\textbf{k}-\textbf{q}} \right. \\ & \left.
        +\frac{4}{3+3\omega}\left(\frac{\Phi^{\prime}_{\textbf{q}}}{\mathcal{H}}+\Phi_{\textbf{q}}\right)\left(\frac{\Phi^{\prime}_{\textbf{k}-\textbf{q}}}{\mathcal{H}}+\Phi_{\textbf{k}-\textbf{q}}\right) \right],
    \end{split}
\end{equation}
the $\omega$ here is the equation of state. The generation of induced gravitational waves is assumed to occur during the radiation-dominated era. Thus, at the time of their generation, we have \cite{Kohri_2018},
\begin{equation}
\small
    \begin{split}
        \Omega_{\text{gw}}(t_f,k) &= \frac{1}{12} \int_0^{\infty} dv \int_{|1-v|}^{1+v} du \left( \frac{4v^2-(1+v^2-u^2)^2}{4uv}\right)^2 \\ & \times \mathcal{P}_{\zeta}(ku) \mathcal{P}_{\zeta}(kv) \left(\frac{3(u^2+v^2-3)}{4u^3v^3} \right)^2 \\ &
        \times \left.\biggr[\left(\pi^2(-3+v^2+u^2)^2 \theta(-\sqrt{3}+u+v) \right) \right. \\ & \left.+\left(-4uv+(v^2+u^2-3)\text{Log} \left| \frac{3-(u+v)^2}{3-(u-v)^2} \right| \right)^2 \right].
    \end{split}
\end{equation}
We can determine the energy density parameter for the current time as follows \cite{Kohri_2018,Ando_2018},
\begin{equation}
    \Omega_{\text{gw}}=\Omega_{rad,0}\Omega_{\text{gw}}(t_f)
\end{equation}
where we have multiplied $\Omega_{\text{gw}}(t_f)$ with the today radiation energy density parameter, $\Omega_{rad,0}$. The GW spectrum for our inﬂation model is shown in Fig.~\ref{fig:gw} along with  the sensitivity curves of the various existing and forthcoming GW experiments.

In order to explain the 15-year data from NanoGrav \cite{Afzal_2023}, it is necessary for the peak in the power spectrum of a Gaussian form to be centered within a specific frequency range, $\log_{10}(f_*/Hz) \sim [-7.25, -5.65]$, with a width denoted as $\Delta$ lying below $2$ at 68\% confidence level \cite{Afzal_2023}. However, our model exhibits a power spectrum with broader width, $\Delta > 4$,  while centered at a higher frequency value, $f_* > 10^{-4}$~Hz. To shift the peak towards lower values of $f$, a discrepancy arises in the spectral index when compared to the Planck data. Consequently, achieving the required Planck normalization while also explaining the NANOGrav observations poses a substantial challenge. This issue is also addressed in \cite{gorji2023extratensorinduced}. Alternatively, see \cite{cheung2023nanograv} for a modified Higgs inflation scenario where the NANOGrav 15-year signal can also be explained with the introduction of a dip in the potential. For a recent model of single-field inflation containing an intermediate phase of ultra-slow-roll see Ref.~\cite{firouzjahi2023induced}.

\begin{figure}[t!]\centering
\includegraphics[width=0.38\textwidth]{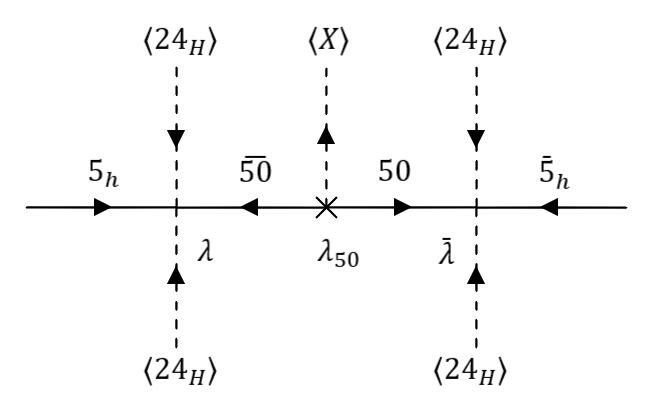}
\caption{\label{5h5hb}Feynman diagram representing mass of color triplets $M_T$.}
\end{figure}
\begin{table}[t!]
\caption{\label{bm} Benchmark point we used for our study.}
\begin{ruledtabular}
\begin{tabular}{llll}
~~~$M$ & $1\times 10^{17}$ GeV &$m_s$ & $10$ TeV ~~~\\
\hline
~~~$M_{GUT}$ & $4\times 10^{16}$ GeV &$g$& 0.7~~~\\
\hline
~~~$M_{50}$ & $1\times 10^{17}$ GeV &$g_5[M_{GUT}]$& 0.78~~~\\
\hline
~~~$M_{\chi}$ & $4\times 10^{17}$ GeV &$\lambda$& $1$~~~\\
\hline
~~~$M_{\Sigma_3}$ & $3\times 10^{14}$ GeV  &$\bar{\lambda}$& $1$~~~\\
\hline
~~~$M_{\Sigma_8}$ & $1\times10^{15}$ GeV  &$\gamma_1$& $-0.44$~~~\\
\hline
~~~$M_{T}$ & $2\times 10^{14}$ GeV &$\gamma_2$& $0.47$~~~ \\
\end{tabular}
\end{ruledtabular}
\end{table}
\section{Gauge Coupling Unification (GCU)\label{gcusec}}
The presence of massless states, such as the color octet $\Sigma_8$ and $SU(2)_L$ triplet $\Sigma_3 $ in the present model, is a generic feature of R-symmetric GUTs \cite{Kyae:2004ft, Fallbacher:2011xg}. Given that $U(1)_R$ is a global symmetry, we can induce intermediate masses for $\Sigma_8$ and $\Sigma_3$ by assuming its breaking at a nonrenormalizable level. This is achieved through the inclusion of the following terms in the superpotential,
\begin{eqnarray}
W_{\slashed{R}}&\supset &
 \gamma_1 \frac{(Tr[\Sigma^2])^2}{M_P}
+\gamma_2 \frac{Tr[\Sigma^4]}{M_P},\\
&\supset &
 M_8\Sigma_8 \Sigma_8
 + M_3 \Sigma_3 \Sigma_3,
\end{eqnarray}
where,
\begin{eqnarray}
M_8 &=& \frac{\gamma_1 }{M_P} \frac{2M^2}{5}+\frac{\gamma_2 }{M_P} M^2, \\
M_3 &=& \frac{\gamma_1 }{M_P} \frac{9 M^2}{10}+\frac{\gamma_2 }{M_P} M^2.
\end{eqnarray}
\begin{figure}[t!]\centering
\includegraphics[width=0.5\textwidth]{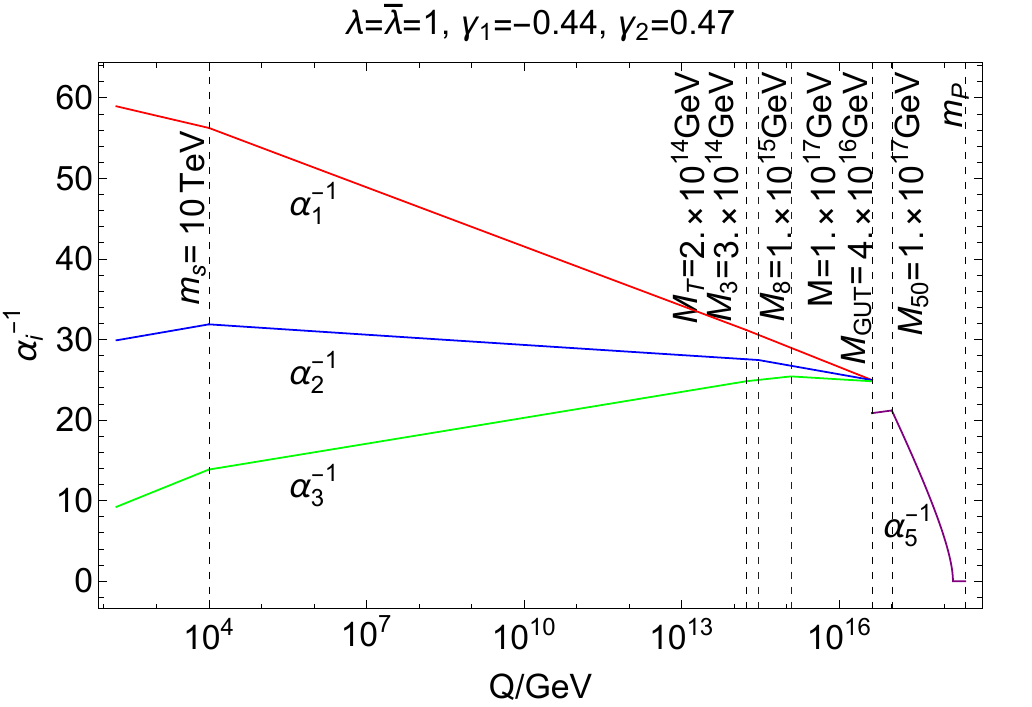}
\caption{\label{gcufig} Two loop evolution of gauge coupling RGEs for $SU(3)_c\times SU(2)_L\times U(1)_Y$ and $SU(5)$.}
\end{figure}
Alternatively, if we do not assume R-symmetry breaking at a nonrenormalizable level, the color octet and $SU(2)_L$ triplet may acquire masses at the order of the TeV scale \cite{Dvali:1997uq} due to soft SUSY breaking terms, thereby ruining the GCU. However, it is possible to restore GCU by introducing additional light vector-like fields \cite{Khalil:2010cp, Rehman:2018gnr, Masoud:2019gxx,Ahmed:2022thr}. These extra light fields can naturally arise in the framework of a missing partner mechanism, as shown in \cite{Antusch:2023mxx}.

Mass matrix for the interactions responsible for masses of color triplets  is,
\begin{eqnarray}\label{mttbar}
\mathcal{M}_{T\overline{T}}=
\begin{pmatrix}
0  & \bar{\lambda} \langle 24_H\rangle ^2 / M_P\\
{\lambda} \langle 24_H\rangle ^2 / M_P & \lambda_{50}\langle X\rangle 
\end{pmatrix},
\end{eqnarray}
this gives eigen value as mass of color triplet,
\begin{eqnarray}\label{mt}
M_T &\simeq & \frac{\lambda \bar{\lambda}}{\lambda_{50}}\frac{\langle 24_H\rangle^4}{M_P^2\langle X\rangle }.
\end{eqnarray}
Planck scale suppressed dimension five term of adjoint Higgs and gauge superfields is,
\begin{eqnarray}
W_{eff}&\supset & \frac{d}{M_P}Tr[24_H \mathcal{W}\mathcal{W}].
\end{eqnarray}
This term can play a decisive role in viability of a model \cite{Tobe:2003yj}.  This dimension five term and masses around or below GUT scale play significant role in GUT scale matching conditions,
\begin{eqnarray}
\frac{1}{g_1^2(Q)}&=&-\frac{8 d M}{M_P}+\frac{1}{g_5^2(Q)}+\frac{1}{8 \pi ^2}\left[\frac{2}{5} \log \left(\frac{Q}{M_{T}}\right)\right.\nonumber \\
&-&\left. 10 \log \left(\frac{Q}{M_X}\right)\right],\\
\frac{1}{g_2^2(Q)}&=&-\frac{3\ 8 d M}{M_P}+\frac{1}{g_5^2(Q)}+\frac{1}{8 \pi ^2}\left[2 \log \left(\frac{Q}{M_{\Sigma3}}\right)\right.\nonumber \\
&-&\left. 6 \log \left(\frac{Q}{M_X}\right)\right],\\
\frac{1}{g_3^2(Q)}&=&\frac{2\ 8 d M}{M_P}+\frac{1}{g_5^2(Q)}+\frac{1}{8 \pi ^2}\left[\log \left(\frac{Q}{M_{T}}\right)\right.\nonumber \\
&+&\left. 3 \log \left(\frac{Q}{M_{\Sigma 8}}\right)-4 \log \left(\frac{Q}{M_X}\right)\right].
\end{eqnarray}
where $M_X = 5g_5 M$ is the mass of gauge bosons.
We define,
\begin{eqnarray}
\epsilon &\equiv & g_3(M_{GUT}) /g,
\end{eqnarray}
where $g=g_1(M_{GUT})=g_2(M_{GUT})$. We can find constraint on masses that is independent of $d$,
\begin{eqnarray}\label{M}
M =\frac{e^{{\pi ^2 \left(1-\frac{1}{\epsilon^2}\right)}/{3 g^2}}{M_P^{1/4}} }{\left(225 {\gamma_1}^2 +812.5 {\gamma_1} {\gamma_2} +625 {\gamma_2}^2 \right)^{1/8}}\frac{M_{GUT}^{3/4}}{g_5^{1/2}},
\end{eqnarray}
and matching condition for gauge couplings of $SU(5)$ and MSSM, for $\lambda=\bar{\lambda}=1$, $\gamma_1 =-0.44$ and $\gamma_2=0.47$, is given as,
\begin{eqnarray}
g_{12} &\simeq & \frac{1}{\sqrt{
g_5^{- 2} +0.34\  + 0.08\ \log [g_5]}}.
\end{eqnarray}

In the present model, several fields, including the color octet $\Sigma_8$, $SU(2)_L$ triplet $\Sigma_3 $, gauge bosons $\chi$ and a pair of color triplets $(T,\bar{T})$ assume a critical role in achieving GCU and facilitating the matching of $SU(5)$ with MSSM.  
For our analysis, we use $M=10^{17}$ GeV, corresponding values for other parameters are given in table \ref{bm}. Two loop renormalization group equations for gauge couplings of standard model gauge group and $SU(5)$ are plotted in Fig. \ref{gcufig}. 
\section{Proton decay\label{pd}}
Proton decay is a very important discriminator to distinguish between different Grand Unified Theories (GUTs). Proton decay has been studied vigorously in $SU(5)$ models. 
In R-symmetric $SU(5)$, proton decays from dimension five and six operators. Dimension five operators mediate via color triplets and dimension six operators mediate via both color triplets and gauge bosons. Proton decay mediated via color triplets in dimension five operators in $SU(5)$ models has been discussed frequently in literature \cite{Hisano:1992jj, Hisano:2013exa, Nagata:2013sba, Ellis:2019fwf, Babu:2020ncc}.

\subsection{Dimension five proton decay}
\begin{figure}[t!]\centering
\subfloat[\label{LLLL}$Q\ Q\ Q\ L$]{\includegraphics[width=1.20in]{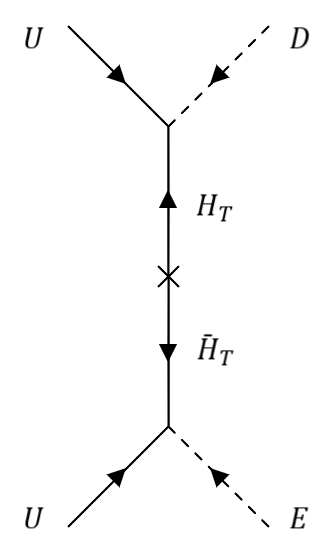}} \quad
\subfloat[\label{RRRR}$U^c D^c U^c E^c$]{\includegraphics[width=1.24in]{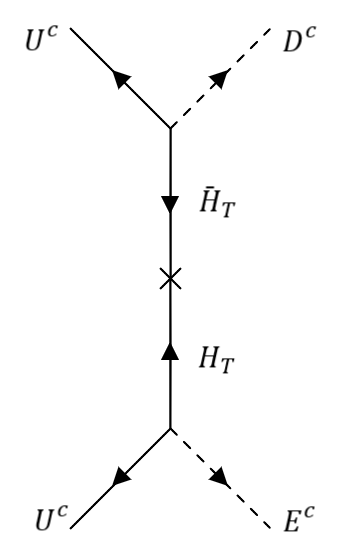}}
\caption{\label{dim5}Chirality flipping, two fermion two scalar, dimension five proton decay operators mediated via color triplets of $5_h,\ \bar{5}_h$.}
\end{figure}

In R-symmetric $SU(5)$ the renormalizable operators involving color triplets $(H_T,\bar{H}_T) \subset (5_h, \bar{5}_h)$ are,
\begin{eqnarray}
y^{(u)} 10\, 10\, 5_h& \supset & y^{(u)} U D H_T+y^{(u)} U^c E^c H_T,\\
y^{(l,d)} 10\, \bar{5}\, \bar{5}_h& \supset & y^{(d)} D N \bar{H}_T + y^{(d)} U^c D^c \bar{H}_T\nonumber \\
&+& y^{(l)} U E \bar{H}_T\, .
\end{eqnarray}
The chirality flipping mass term, $5_h\bar{5}_h$, and Yukawa interactions lead to two scalars and two fermions, dimension five proton decay operators  as shown in Fig.~\ref{dim5}.
The $5_h\bar{5}_h$ term, leads to fast proton decay even with color triplet masses of order GUT scale, but in our R-symmetric $SU(5)$ model, $U(1)_A$ symmetry forbids this term upto all orders.  However, it can arise indirectly from $U(1)_A$ symmetry-breaking terms via the vev $\langle X\rangle$, as shown in Fig. \ref{5h5hb}. 

The chirality flipping dimension five proton decay operator contribute in a loop diagram to effectively become four-Fermi operator. The decay rate, aside from loop factors, varies as,
\begin{eqnarray}
\Gamma &\propto & \frac{\mu}{M_T}\frac{1}{M_T^2},
\end{eqnarray}
where $\mu\sim 100$ GeV, comes from the $\mu$-term generated by GM mechanism.
For typical values of parameters we obtain $M_T \sim 10^{14}$ GeV, and $\mu/M_T \sim 10^{-12}$. Thus we conclude that dimension five proton decay in R-symmetric $SU(5)$ via chirality flipping operator is safe from rapid proton decay. 

\begin{figure}[t!]\centering
\includegraphics[width=0.4\textwidth]{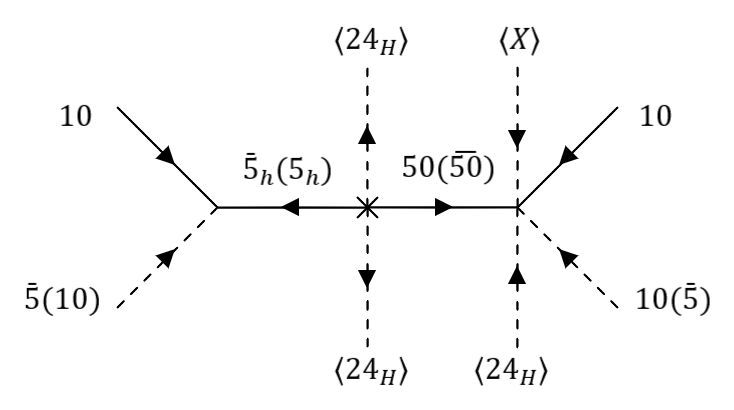}
\caption{\label{hetro}Feynman diagram for dimension five proton decay operators mediated via heterogenous mediators ($5$ and $50$-plets).}
\end{figure}

We studied a chirality flipping dimension five proton decay operator, that arise from intermixing of renormalizable and nonrenormalizable interactions from $W$ as follows,
\begin{eqnarray}
W &\supset & \frac{24_H}{M_P}\frac{\langle X \rangle}{M_P}\left(\eta_1\ 10\ 10\ 50 + \eta_2\  \overline{50}\ 10\ \bar{5}\right).
\end{eqnarray}
These ``\textit{heterogenous}" dimension five operators \cite{Mehmood:2023gmm} involve color triplets coming from two different representations, $5$ and $50$-plets. Feynman diagram for these operators is given in Fig. \ref{hetro}. These operators do not contribute significantly to decay rate of proton due to an extra suppression factor of $(M\langle X\rangle /M_P^2)\lesssim 10^{-3}$. 
\begin{figure}[t!]\centering
\subfloat[$(Q\ U^{c\dagger})\ (Q\ E^{c\dagger})$]{\includegraphics[width=1.40in]{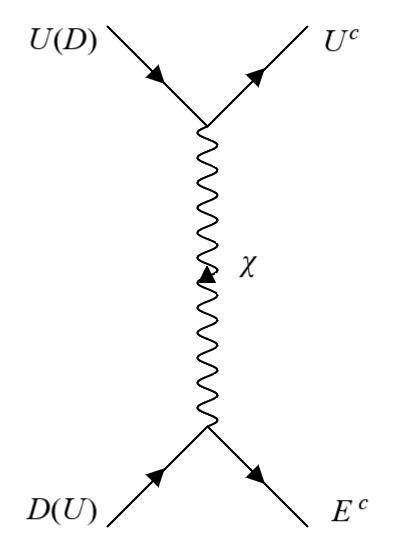}} \quad
\subfloat[$(Q\ U^{c\dagger})\ (L\ D^{c\dagger})$]{\includegraphics[width=1.4in]{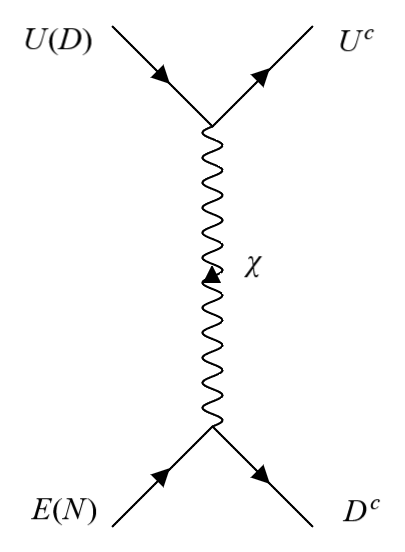}}
\caption{\label{gb}Four-fermion,  dimension six proton decay operators mediated via gauge boson $24_A\supset \chi$.}
\end{figure}
\subsection{Dimension six proton decay}
Dimension six proton decay operators are mediated by the gauge bosons  $(\chi, \bar{\chi}) \subset 24_A$ and the color triplets $(H_T,\bar{H}_T) \subset (5_h, \bar{5}_h)$. 
The interaction terms representing the mediation of gauge bosons  arise from the K\"ahler potential. Feynman diagrams for gauge boson mediated proton decay are shown in Fig.~\ref{gb}.
\begin{figure*}[ht!]\centering
\subfloat[\label{2f2s1}$Q\ Q\ (U^cE^c)^{\dagger}$]{\includegraphics[width=1.30in]{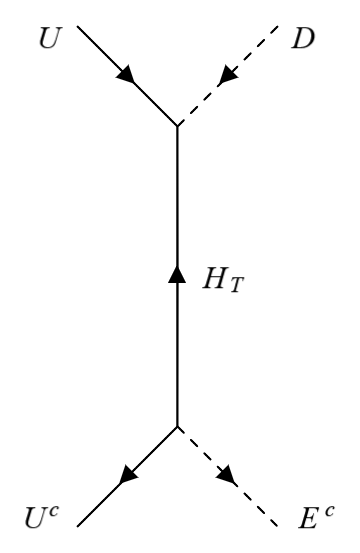}}\quad
\subfloat[\label{2f2s2}$Q\ L\ (U^c D^c)^{\dagger}$]{\includegraphics[width=1.5in]{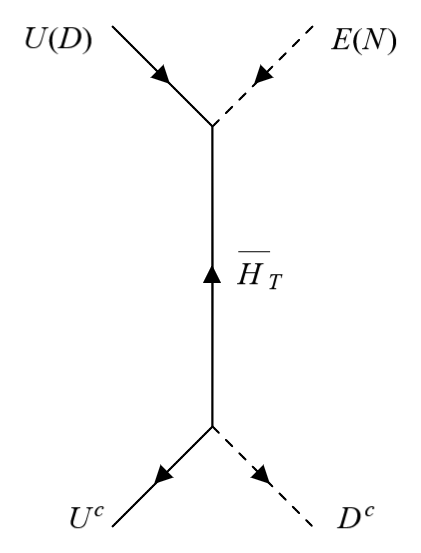}}
\subfloat[\label{4f1}$Q\ Q\ (U^cE^c)^{\dagger}$]{\includegraphics[width=1.3in]{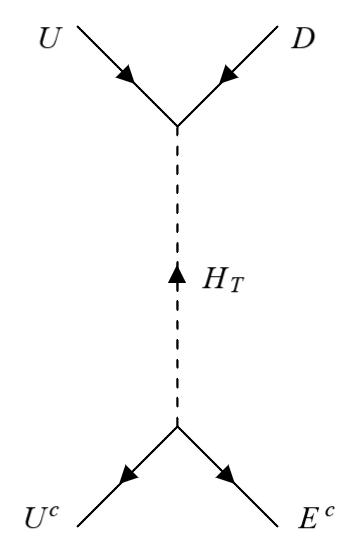}}\quad
\subfloat[\label{4f2}$Q\ L\ (U^c D^c)^{\dagger}$]{\includegraphics[width=1.5in]{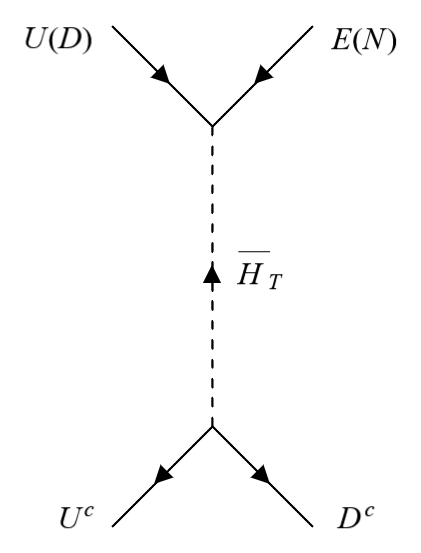}}
\caption{\label{dim6ct}$LLRR$ type dimension six proton decay operators mediated via color triplets of $5_h,\ \bar{5}_h$.}
\end{figure*}
These interactions correspond to proton decay operators of the $LLRR$ chirality type.

Dimension six proton decay operators of chirality type $LLRR$, mediated via color triplets are shown in Feynman diagrams in Fig.~\ref{dim6ct}. 
Two fermions and two scalars diagrams in Figs.~\ref{2f2s1} and \ref{2f2s2} are also dimension six operators, as clarified in \cite{Mehmood:2020irm}.
These operators need a box diagram to effectively become the four-Fermi proton decay operators. 
Hence, loop factors will suppress the contribution of these diagrams as compared to four-fermion diagrams shown in Figs.~\ref{4f1} and \ref{4f2}. 
Thus, only the four-fermion diagrams of Figs.~\ref{4f1} and \ref{4f2} will be considered for chirality nonflipping color-triplet mediation of type $LLRR$.

For chirality nonflipping dimension six operators, mass of color triplets are given in eq. (\ref{mt}),
where we assume couplings $\bar{\lambda}= {\lambda}$ to be real. For $\lambda_{50}=\bar{\lambda}_{50}=1$, $\langle X \rangle =M_{50} \sim 10^{17}\ \text{GeV}$ and $M \sim  10^{17}\ \text{GeV}$ we get,
\begin{eqnarray}
M_T= M_{\bar{T}} 
\sim 
{\lambda}^2 \,( 1.7  \times 10^{14} \text{GeV}).
\end{eqnarray}
In the end, mediation of chirality nonflipping via color triplets and gauge bosons will contribute dominantly in the final decay rates.

\begin{table*}[t!]
\caption{Values of relevant matrix elements, The Super-K bounds, Hyper-K, and DUNE sensitivities for various proton decay channels.}\label{matrix-el}
\begin{ruledtabular}
\begin{tabular}{llllll}
\textrm{Decay channel}&
\multicolumn{1}{c}{\textrm{$T_{\text{ml}}=$Matrix element ($\text{GeV}^2$)}}&
\textrm{Super-K bound \cite{ParticleDataGroup:2018ovx}}&
\multicolumn{2}{c}{\textrm{Sensitivities ($10^{34}$ years)}}\\
&\textrm{}&\textrm{($10^{34}$ years)}&\textrm{Hyper-K \cite{Hyper-Kamiokande:2018ofw}}&\textrm{DUNE \cite{DUNE:2020ypp,DUNE:2015lol}}\\
\colrule
$e^+ \,\pi^0$ &$T_{\pi^0 e^+}=\langle \pi^0|(ud)_R u_L|p\rangle_{e^+}=-0.131(4)(13)$&$~~~~~1.6$ &$~~~~~7.8$&$~~~~~--$\\
\hline
$\mu^+ \, \pi^0$&$T_{\pi^0 \mu^+}=\langle \pi^0|(ud)_R u_L|p\rangle_{\mu^+}= -0.118(3)(12)$&$~~~~~0.77$ &$~~~~~7.7$&$~~~~~--$\\
\hline
$\bar{\nu}_i \, K^+ $&$T^{\prime}_{ \bar{\nu}_i K^+}=\langle K^+|(ud)_R s_L|p\rangle=-0.134(4)(14)$&$~~~~~0.59$&$~~~~~3.2$&$~~~~~6.2$\\
&$T^{\prime \prime}_{\bar{\nu}_i K^+}=\langle K^+|(us)_R d_L|p\rangle=-0.049(2)(5)$&&&\\
\hline
$\bar{\nu}_i \, \pi^+ $&$T_{\bar{\nu}_i \pi^+ }=\langle \pi^+|(ud)_R d_L|p\rangle=-0.186(6)(18)$&$~~~~~0.039$&$~~~~~--$&$~~~~~--$\\
\hline
$e^+ \, K^0$&$T_{K^0 e^+}=\langle K^0|(us)_R u_L|p\rangle_{e^+}=\ 0.103(3)(11)$& $~~~~~0.1$ &$~~~~~--$&$~~~~~--$\\
\hline
$\mu^+ \, K^0$&$T_{K^0 \mu^+}=\langle K^0|(us)_R u_L|p\rangle_{\mu^+}=\ 0.099(2)(10)$&$~~~~~0.16$&$~~~~~--$&$~~~~~--$\\
\end{tabular}
\end{ruledtabular}
\end{table*}
The decay rates for charged lepton channels are,
\begin{widetext}
\begin{eqnarray}
\Gamma_{l^+ \pi^0} &=& 
\mathit{F}_{\pi^0}T^2_{l^+ \pi^0}\left(A_{S_1}^2 \left|\frac{g^2_5}{M^2_{\chi}} V_{1p} +\frac{1}{M^2_{\bar{T}}}Y^{( d)2}_{11} V_{1p} \right|^2 
+ A_{S_2}^2\left|\frac{g^2_5}{M^2_{\chi}}V^*_{1p}+\frac{1}{M^2_{{T}}}Y^{(u)}_{11} Y^{(u)}_{pp} V^{*}_{p1}\right|^2  \right) , \\
\Gamma_{l^+ K^0} &=&
\mathit{F}_{K^0}T^2_{l^+ K^0} \left( A_{S_1}^2\left|\frac{g^2_5}{M^2_{\chi}}V_{2p} +\frac{1}{M^2_{\bar{T}}}Y^{( d)}_{11}Y^{( d)}_{mm}V_{21}V^*_{1m}V_{mp}\ \right|^2 
+A_{S_2}^2\left|\frac{g^2_5}{M^2_{\chi}}V^{*}_{2p}\right|^2\right) ,
\end{eqnarray}
\end{widetext}
where,
\begin{eqnarray}
\mathit{F}_x &=& \frac{m_p A^2_L}{32\pi}\left(1-\frac{m^2_{x}}{m^2_p}\right)^2,
\end{eqnarray}
\begin{figure*}[t!]\centering
\subfloat[\label{MTvsM}]{\includegraphics[width=0.47\textwidth]{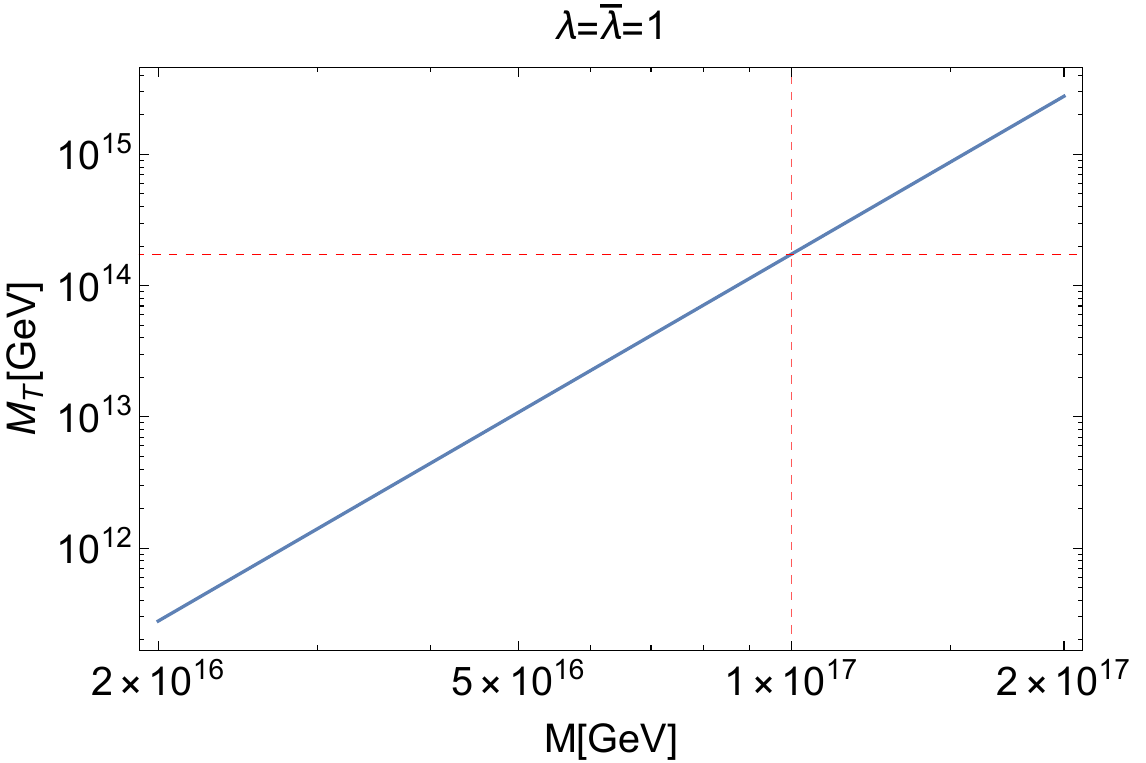}}\qquad
\subfloat[\label{gammavsM}]{\includegraphics[width=0.47\textwidth]{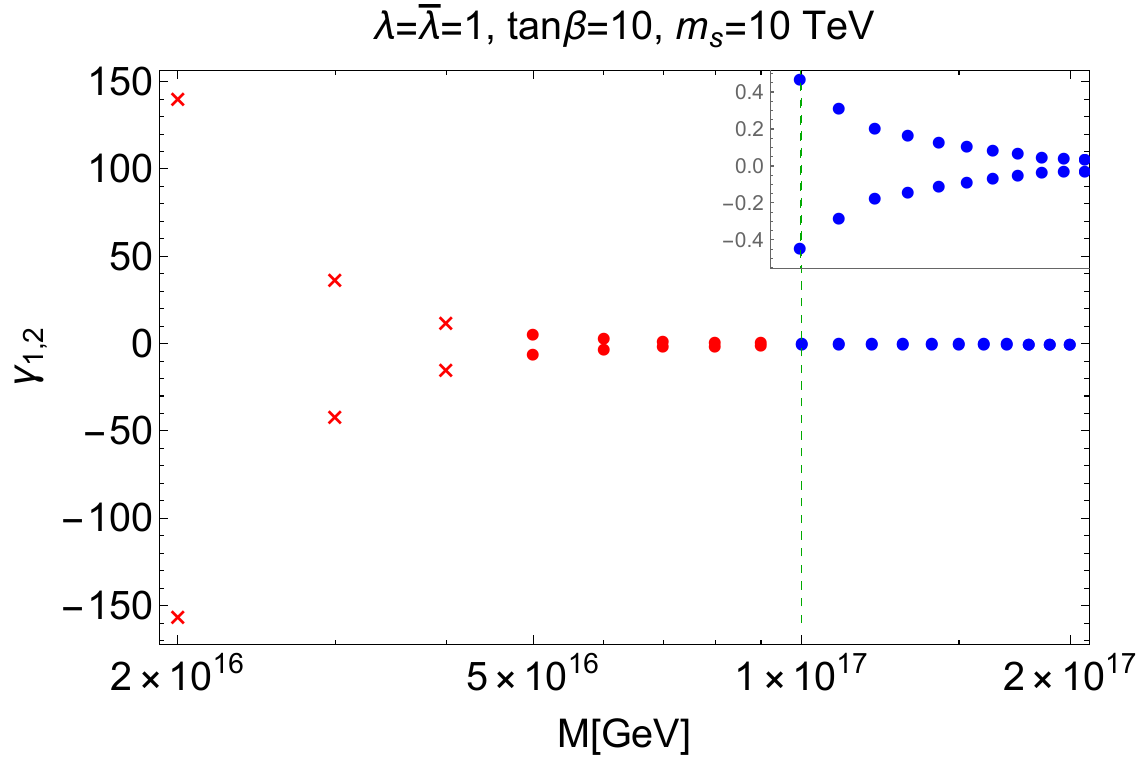}}
\caption{ (a) shows variation of color triplet mass $M_T$ vs. $M$ for fixed value of $\lambda$. Red (dashed) lines represent corresponding values for our benchmark point. (b) shows variation of couplings $\gamma_1$ and $\gamma_2$ with respect to $M$. Plot markers `$\times$' (`${\bullet}$') show the values forbidden (allowed) by proton decay bound for $\bar{\nu}K^+$ channel for fixed values of $\lambda$, $\tan \beta$ and $m_s$.}
\end{figure*}
\begin{figure*}[ht!]\centering
\subfloat[$p \rightarrow e^+ \pi^0$]{\includegraphics[width=0.45\textwidth]{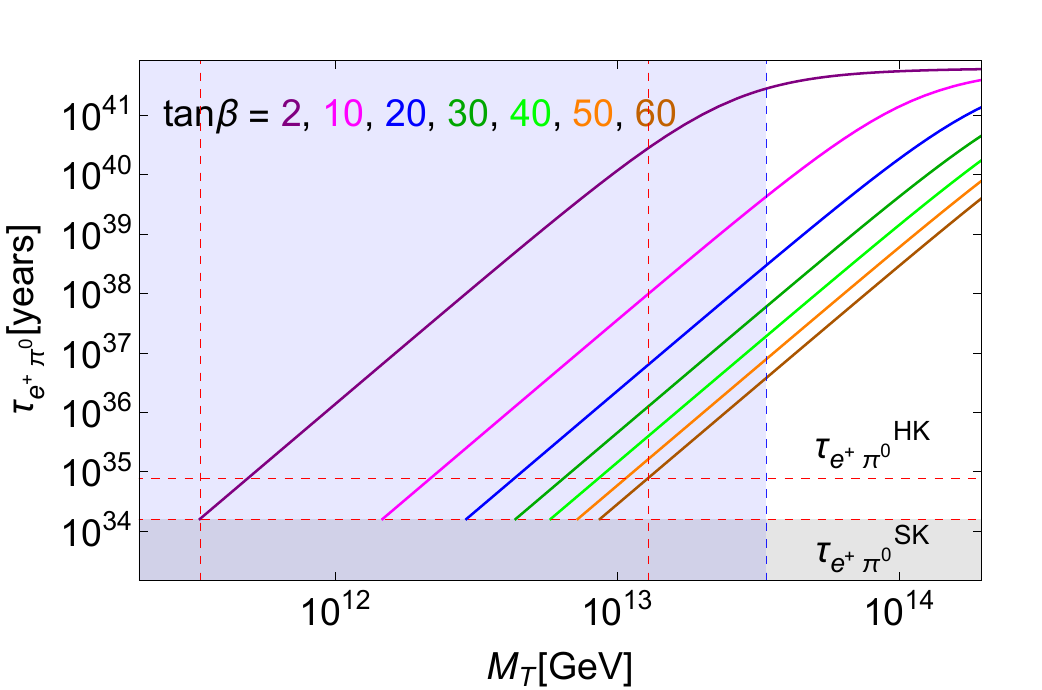}}
\subfloat[$p \rightarrow \mu^+ \pi^0$]{\includegraphics[width=0.45\textwidth]{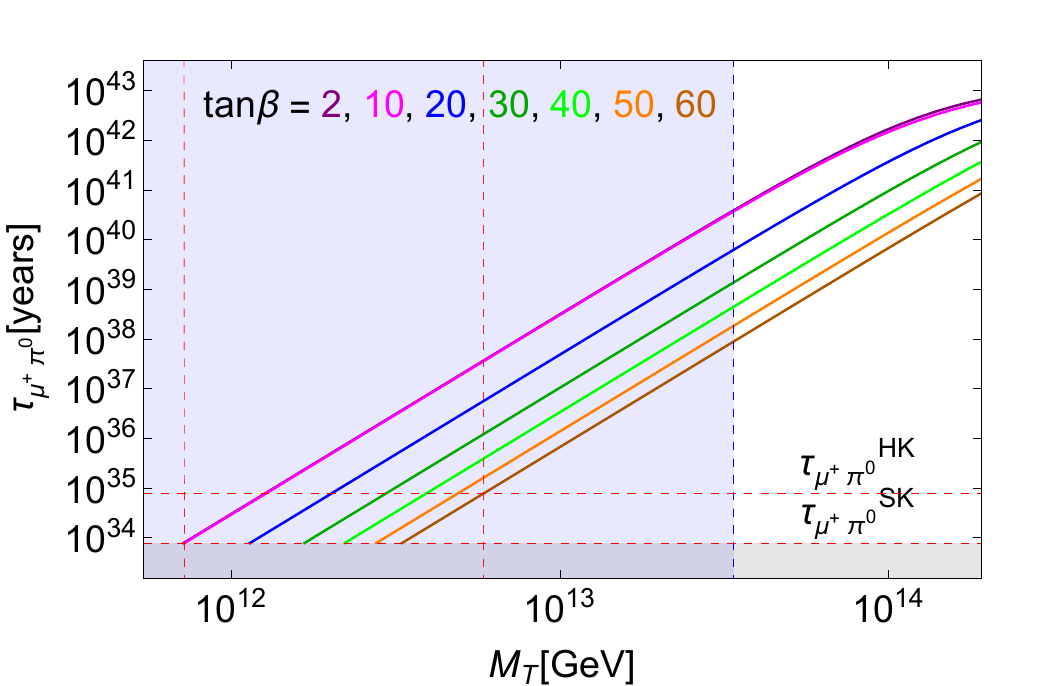}}\\
\subfloat[$p \rightarrow e^+ K^0$]{\includegraphics[width=0.45\textwidth]{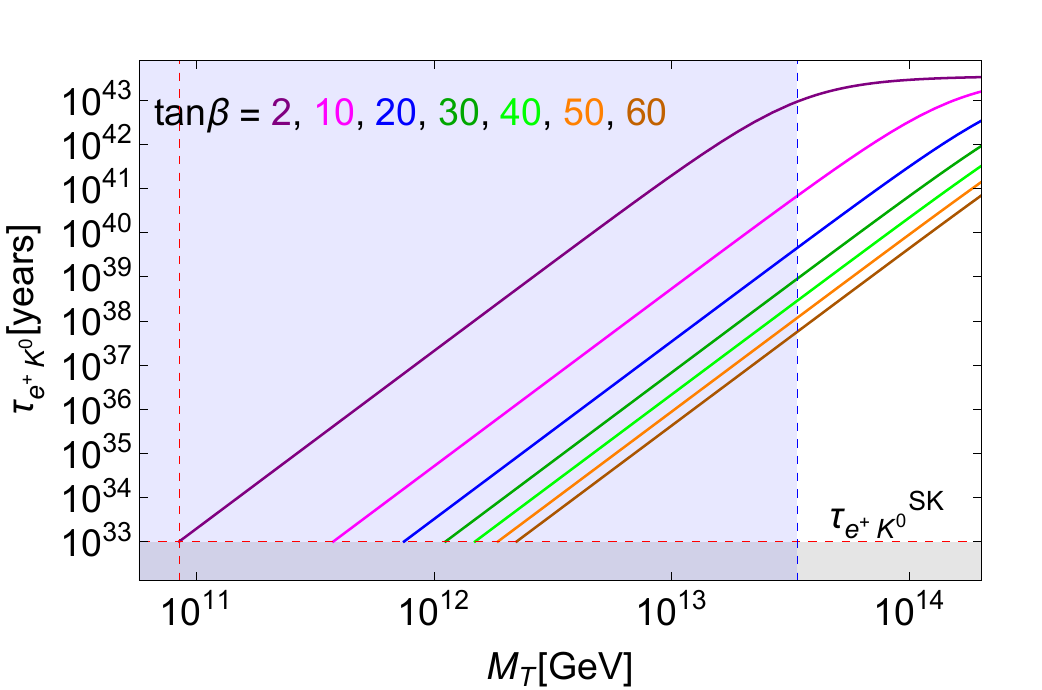}}
\subfloat[$p \rightarrow \mu^+ K^0$]{\includegraphics[width=0.45\textwidth]{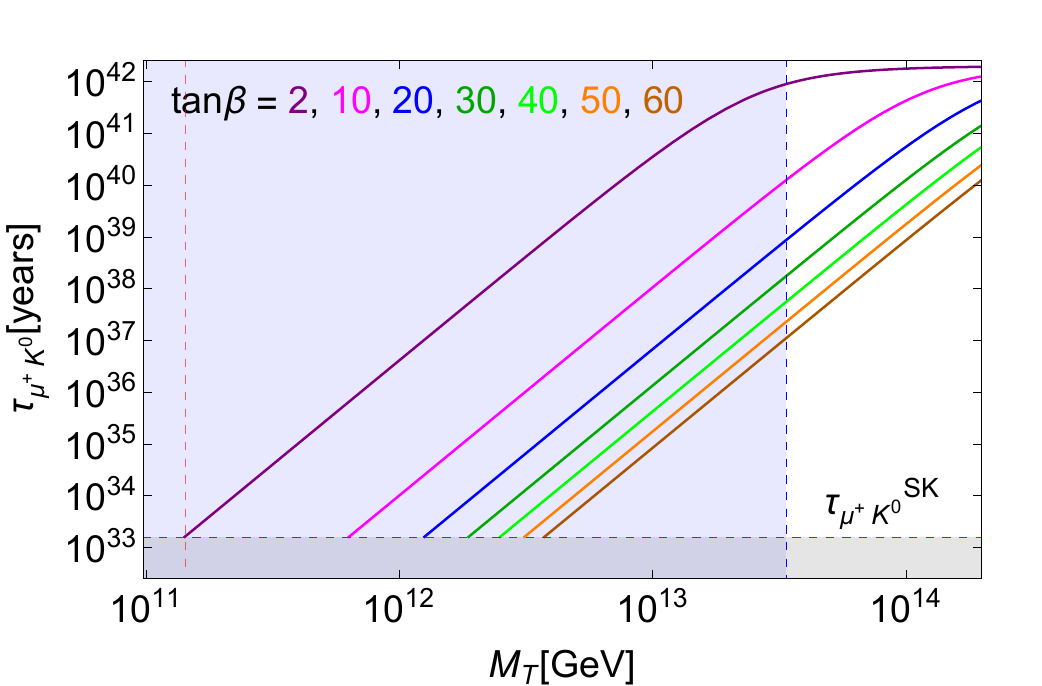}}
\caption{\label{clpd} The partial-lifetime estimates of proton for charged-lepton decay channels as a function of  $M_T $ with $\tan \beta$ in the range, $2\leq\tan{\beta}\leq 60$ (increasing from left to right). The bottom red (dashed) lines represent the experimental limits from Super-K and the top red (dashed) lines represent expected sensitivity of Hyper-K in future. }
\end{figure*}
and  $Y^{(x)}$ represents diagonalized Yukawa couplings, $V$ represents the CKM matrix, $g_5$ is $SU(5)$ gauge coupling at GUT scale and $T_{lm}$ represents hadronic marix elements given in table \ref{matrix-el}\,\cite{Aoki:2017puj}. Current experimental bounds from Super-K and future sensitivities of Hyper-K and DUNE on different proton decay channels  are  given in table \ref{matrix-el}. 

The decay rates for the neutral lepton channels are,
\begin{widetext}
\begin{eqnarray}
\Gamma_{ \bar{\nu}_i K^+ }
&=&\mathit{F}_{K^+}A_{S_1}^2\left|\frac{g^2_5T''_{ \bar{\nu}_i K^+}}{M^2_{\chi}}\delta_{2i}
+\frac{1}{M^2_{\bar{T}}}Y^{( d)}_{11}Y^{( d)}_{ii}\left(T''_{ \bar{\nu}_i K^+ }V_{21}V^*_{1i}
+ T'_{ \bar{\nu}_i K^+}V_{11}V^*_{2i}\right)\right|^2 \, ,\\
\Gamma_{ \bar{\nu}_i \pi^+} &=& \mathit{F}_{\pi^+}A_{S_1}^2T^2_{ \bar{\nu}_i \pi^+ }
\left|\frac{g^2_5}{M^2_{\chi}} +\frac{Y^{( d)\,2}_{11}}{M^2_{\bar{T}}}\ \right|^2\, .
\end{eqnarray}
\end{widetext}

 Lifetime of proton depends on masses of mediators (color triplets and gauge bosons) which depend upon value of M. In Fig. \ref{MTvsM}, mass of color triplet $M_T$ is plotted w.r.t. $M$ for $\lambda = \bar{\lambda}=1$.
This shows that proton decay analysis crucially depends on value of $M$. In Fig. \ref{gammavsM}, we see variation of $\gamma_1 <0$ and $\gamma_2>0$ w.r.t. $M$ for $\lambda = \bar{\lambda}=1$, $\tan \beta =10$ and $m_s=10$ TeV.
We assumed degeneracy of all supersymmetric particles at $m_s$. In this plot, `$\times$' (`${\bullet}$') shaped plot markers represents that these values are forbidden (allowed) by proton decay bound set by Super-K for $\bar{\nu}_iK^+$ channel. Red (Blue) colored points represents that $\gamma_2>1$ ($\gamma_2<1$). It is to be noted that $\gamma_2<1$ for $M\geq 10^{17}$ GeV, hence for our analysis we are using  $M\sim 10^{17}$ GeV. 

For numerical estimates we assume $\lambda = \bar{\lambda} $.
The plots in Fig.~\ref{clpd} shows the estimated partial lifetime of proton for charged lepton channels as a function of color triplet mass $M_T$ for values of  $\tan \beta$ ranging from $2$ to $60$.
The lower horizontal red line in each plot represents the Super-K bounds on lifetime of proton, while the upper horizontal red line represent the expected sensitivities of future experiments. 
The gray shaded region in each plot represents the range of proton lifetimes that have been excluded by Super-K.
\begin{figure*}[t!]\centering
\subfloat[\label{dune}$p \rightarrow \bar{\nu}_i K^+$]{\includegraphics[width=0.45\textwidth]{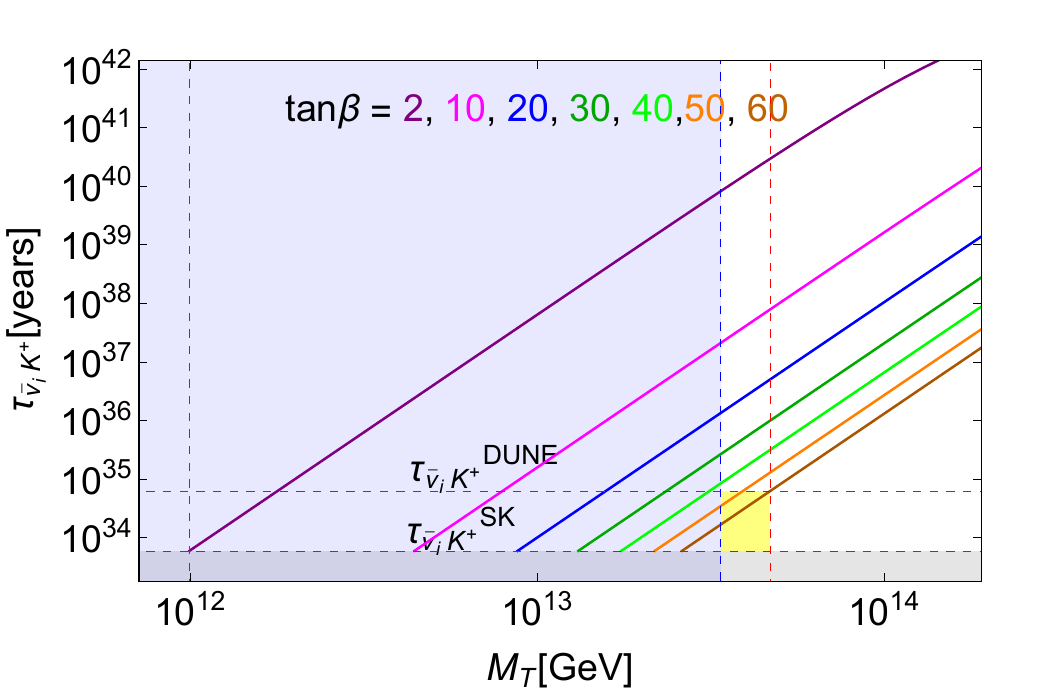}}\qquad
\subfloat[$p \rightarrow \bar{\nu}_i \pi^+$]{\includegraphics[width=0.45\textwidth]{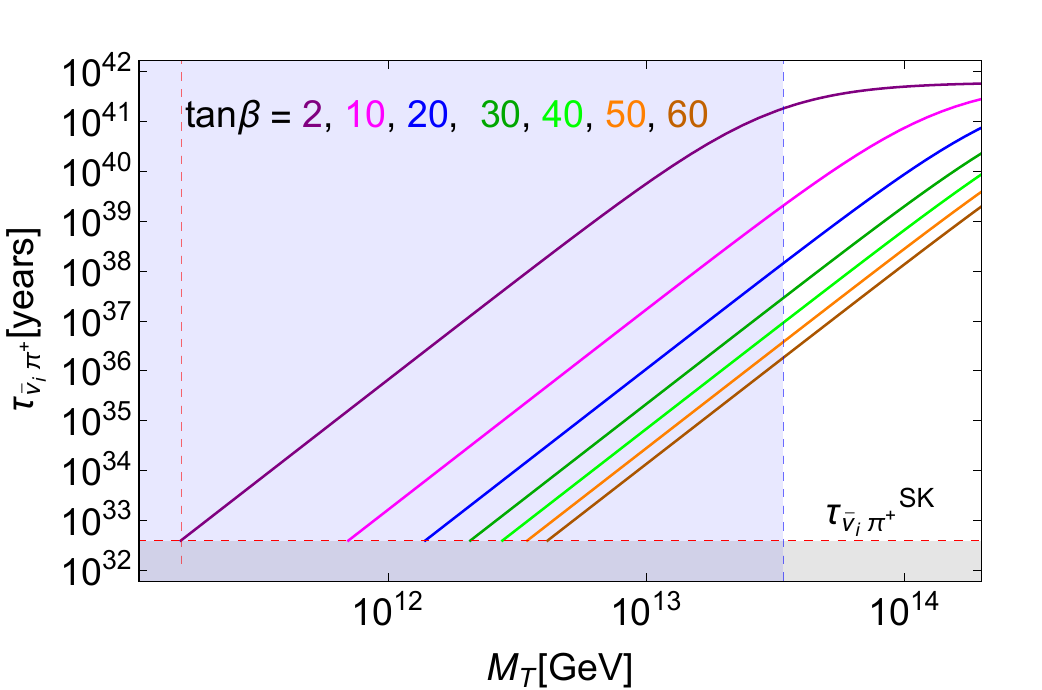}}
\caption{\label{nlpd} The partial-lifetime estimates of proton for neutral lepton decay channels as a function of $M_T$  with $\tan \beta$ in the range $2\leq\tan{\beta}\leq 60$ (increasing from left to right). The bottom red (dashed) line represents the experimental limit from Super-K and thetop red (dashed) line in (b)) represents the expected sensitivity of DUNE in future.}
\end{figure*}
\begin{figure*}[t!]\centering
\subfloat[\label{kvsM}]{\includegraphics[width=0.45\textwidth]{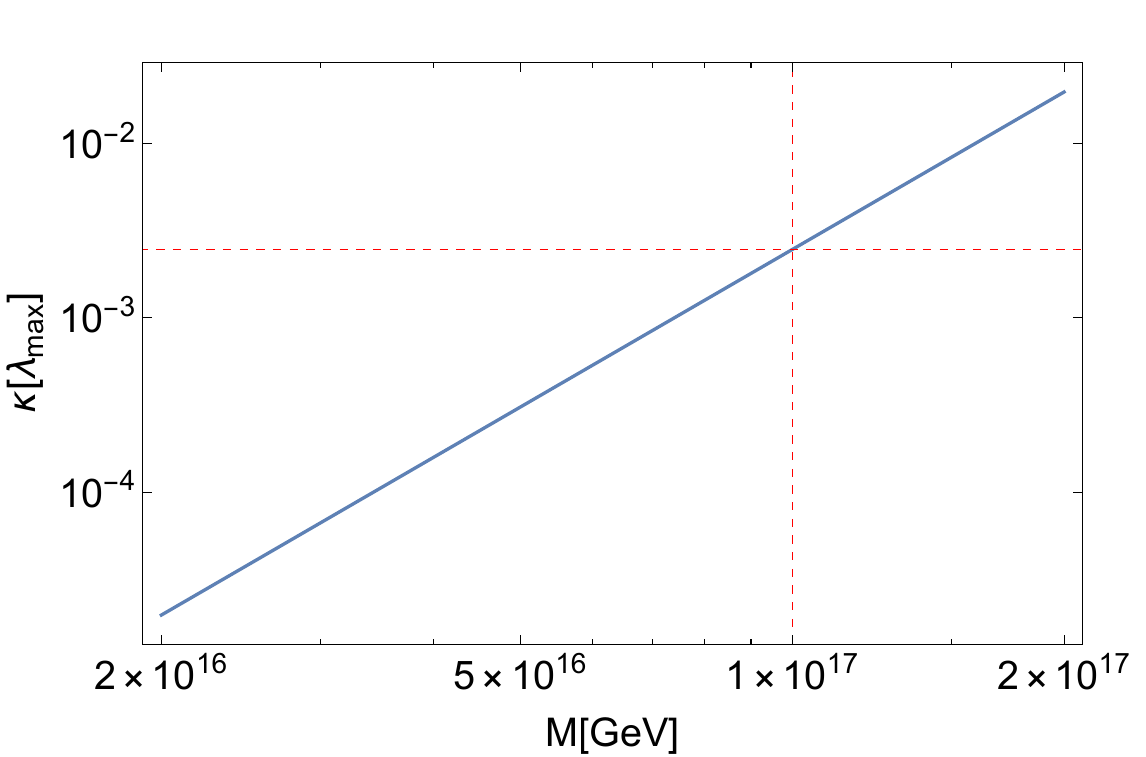}}\qquad
\subfloat[\label{lamvsk}]{\includegraphics[width=0.45\textwidth]{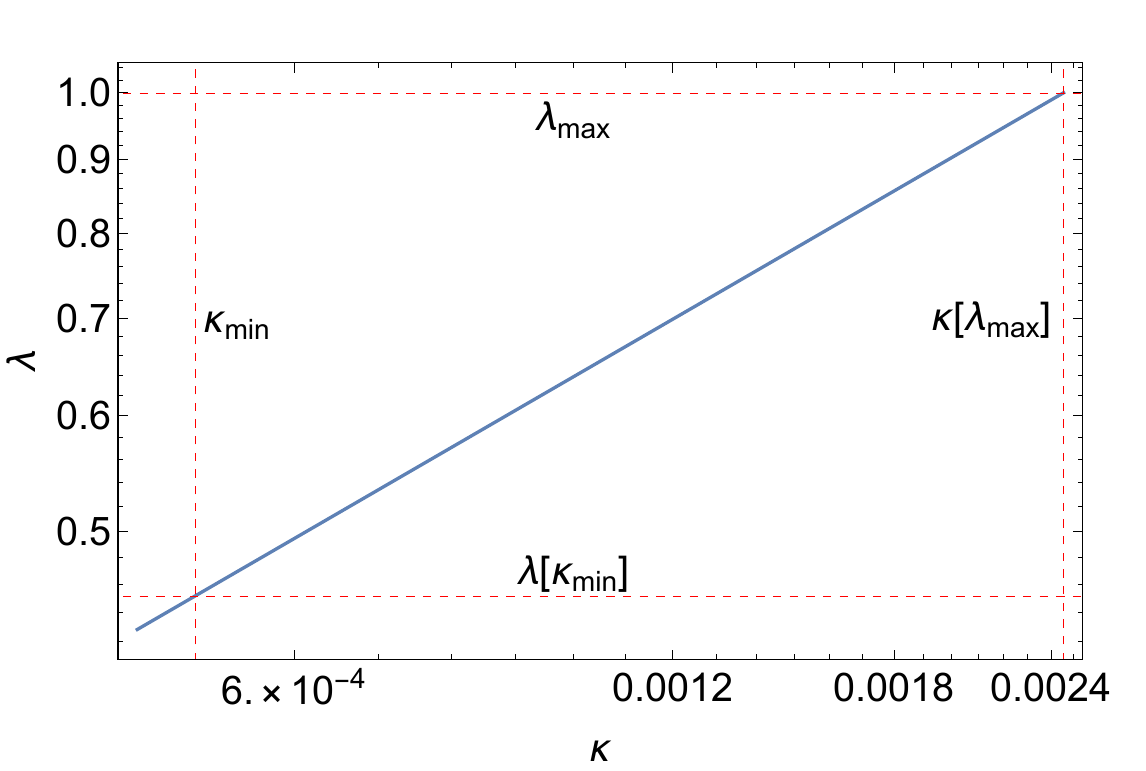}}
\caption{Fig. (a) shows variation of $\kappa[\lambda_{\text{max}}]$ vs. $M$ . Intersection of red (dashed) lines corresponds to benchmark point of our study. Fig. (b) shows variation of $\lambda$ vs. $\kappa$ for a fixed $M\sim 10^{17}$ Gev. Upper horizontal red (dashed) line corresponds to maximum value of coupling of $\lambda$ taken for proton decay study. Left vertical red (dashed) line corresponds to minimum value of $\kappa$ set by inflation.}
\end{figure*}
\begin{figure*}[t!]\centering
\subfloat[\label{lam}]{\includegraphics[width=0.49\textwidth]{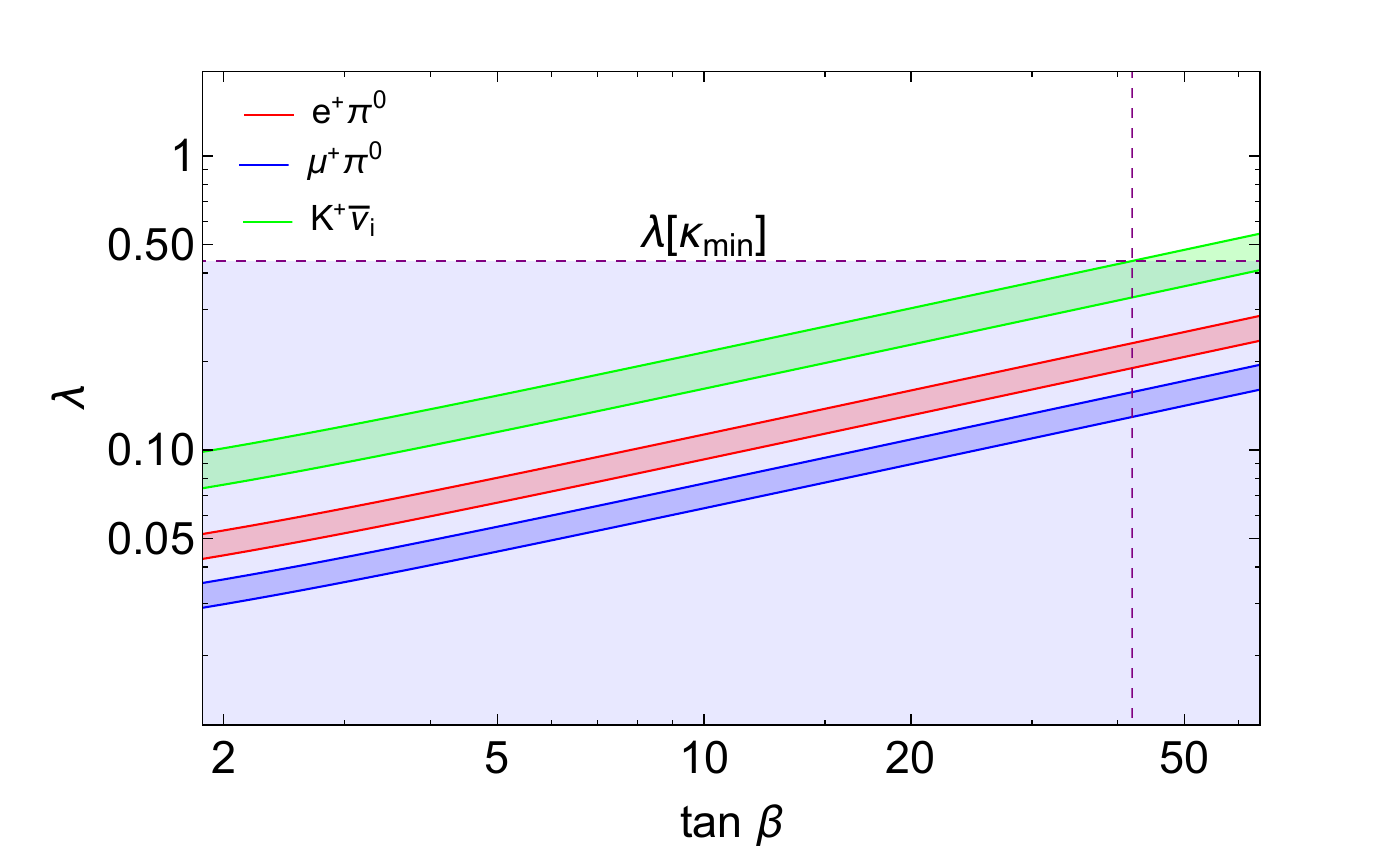}}
\subfloat[\label{Mt}]{\includegraphics[width=0.49\textwidth]{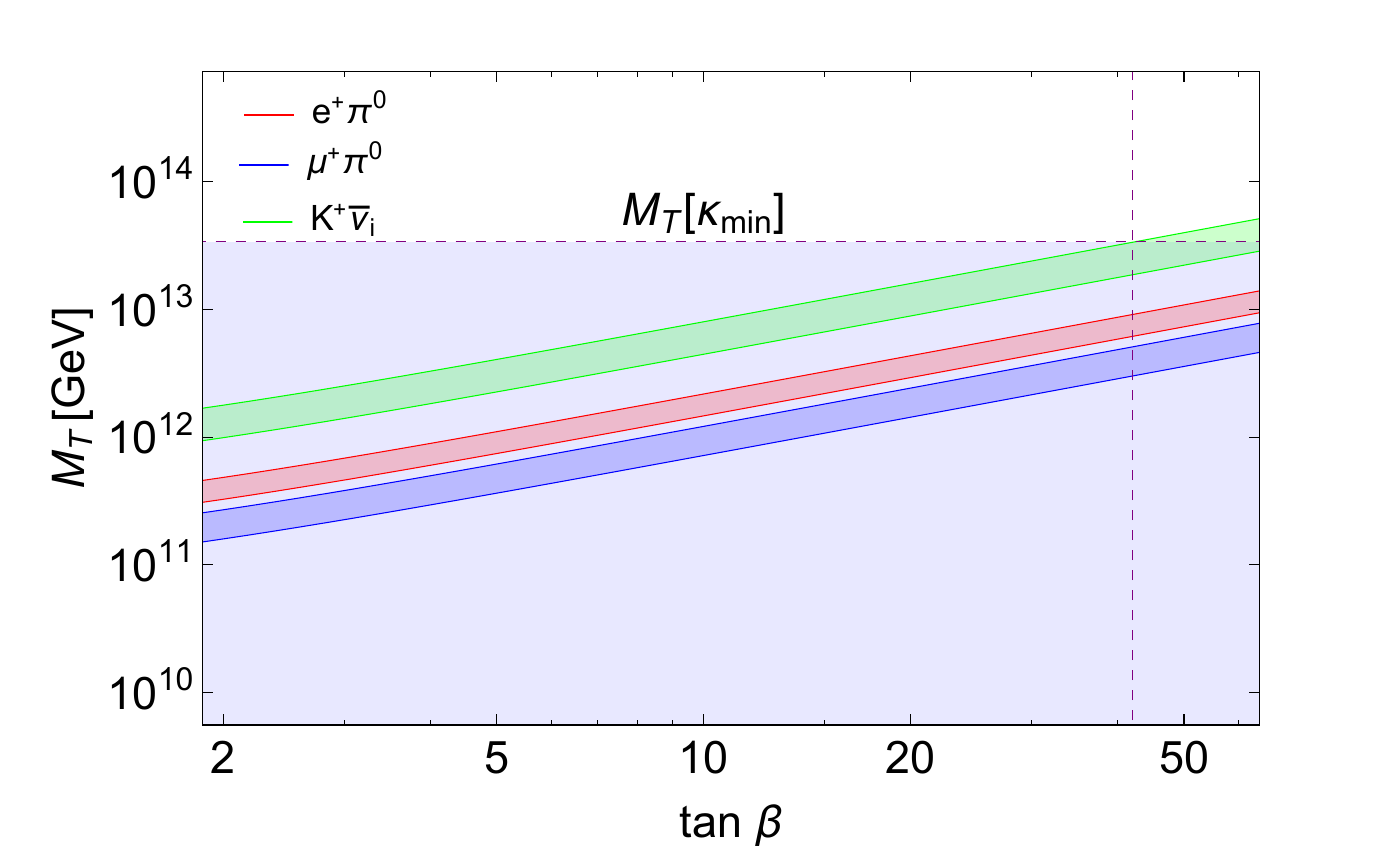}}\
\caption{\label{vstb}Variation of range of $\lambda$ and $M_T$ for different proton decy channels set by Super-K and future experiments in \ref{lam} and\ref{Mt} vs. $\tan \beta$. Shaded region is excluded by the bound set by $\kappa=0.0005$.}
\end{figure*} 
\begin{figure*}[t!]\centering
\subfloat[$\Gamma_{\mu^+\pi^0}/\Gamma_{e^+\pi^0}$]{\includegraphics[width=0.33\textwidth]{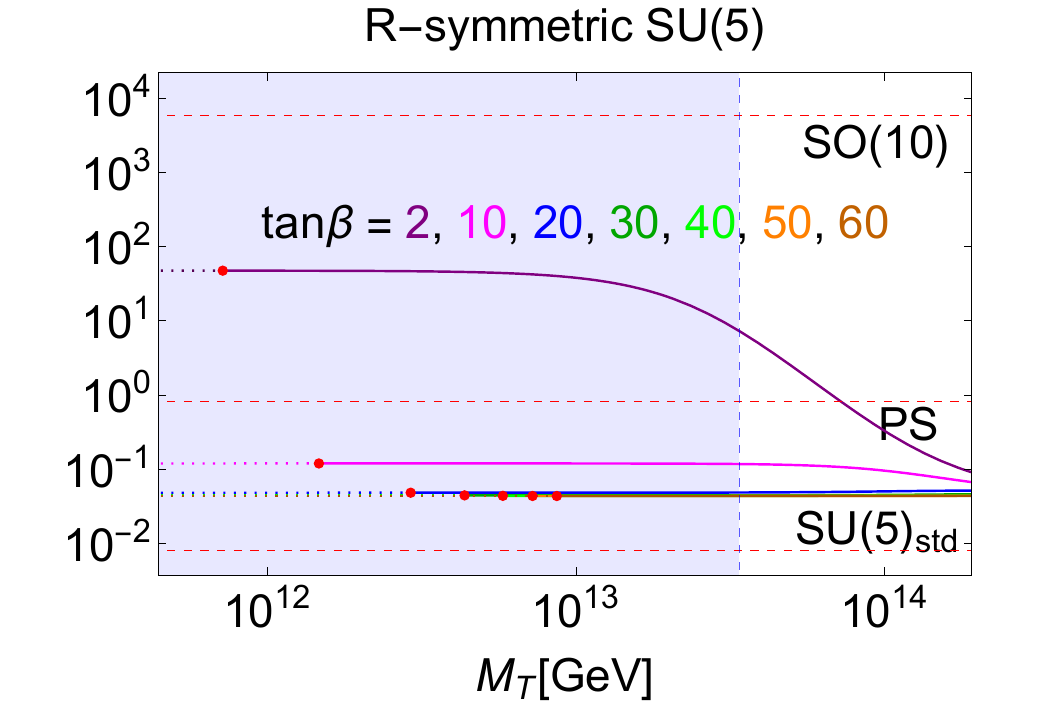}
\includegraphics[width=0.33\textwidth]{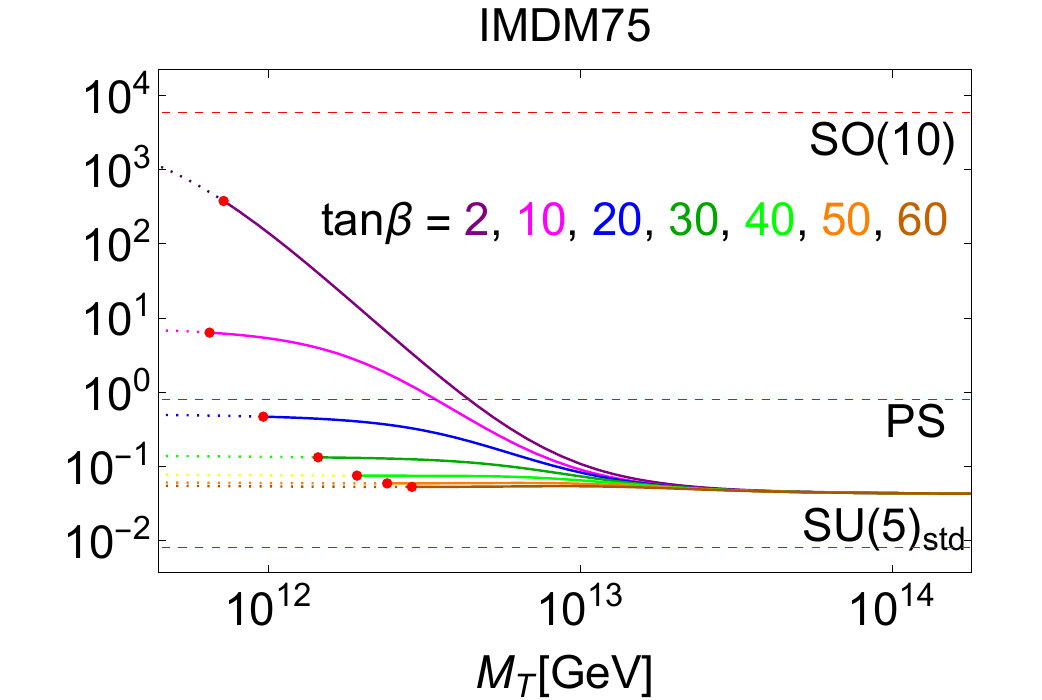}
\includegraphics[width=0.33\textwidth]{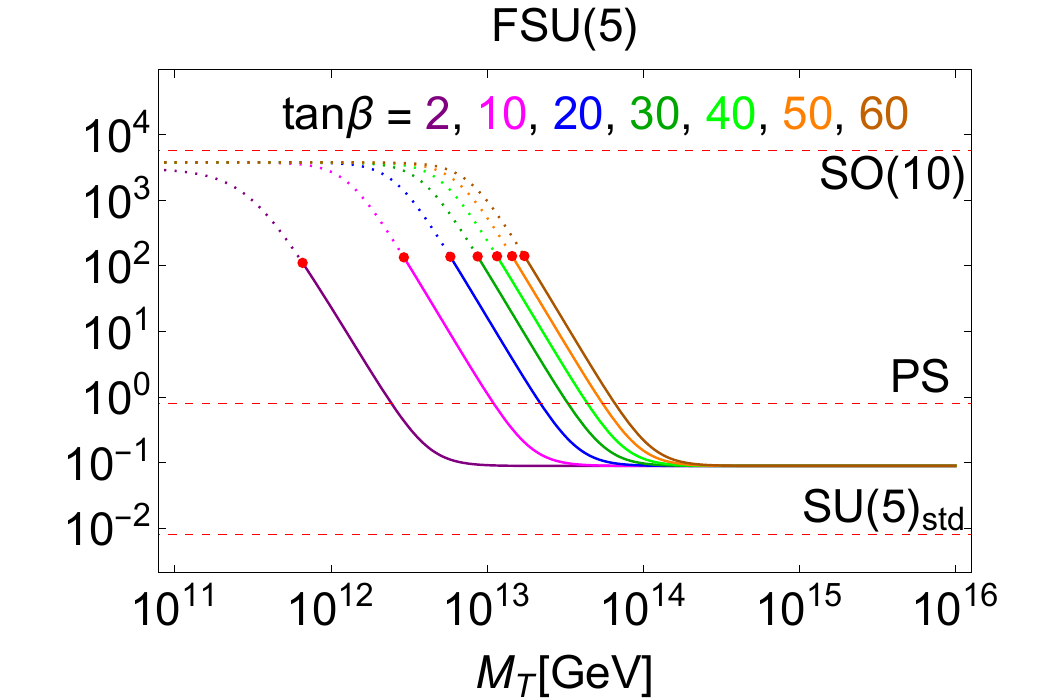}}\\
\subfloat[\label{nupiepi} $\Gamma_{\bar{\nu}_i\pi^+}/\Gamma_{e^+\pi^0}$]{\includegraphics[width=0.33\textwidth]{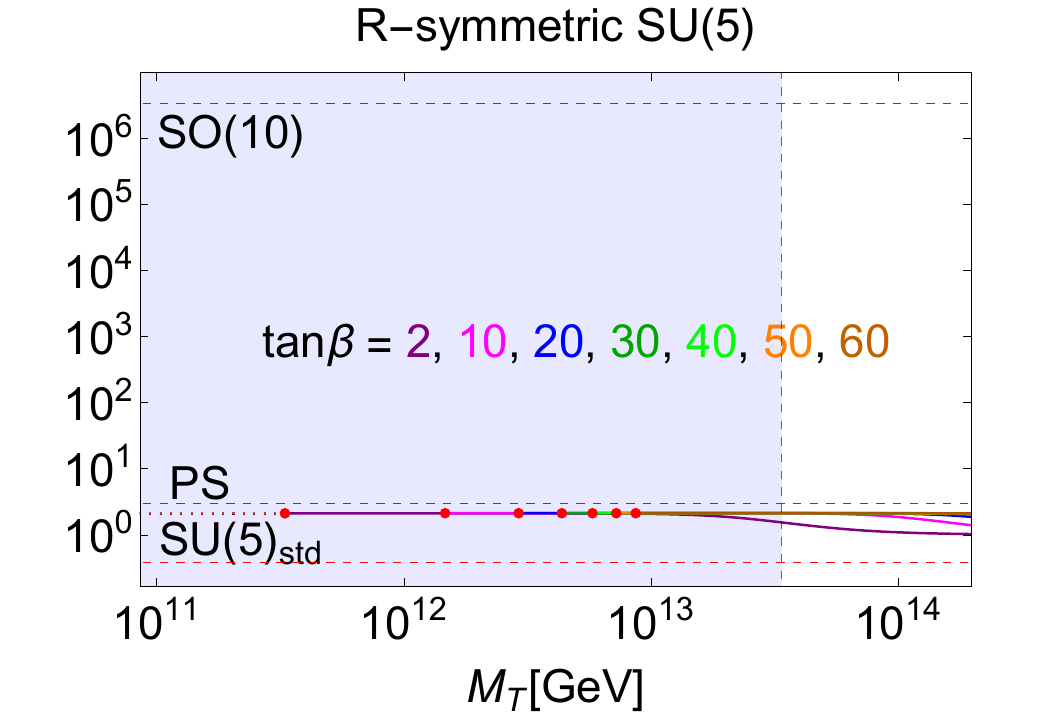}\includegraphics[width=0.33\textwidth]{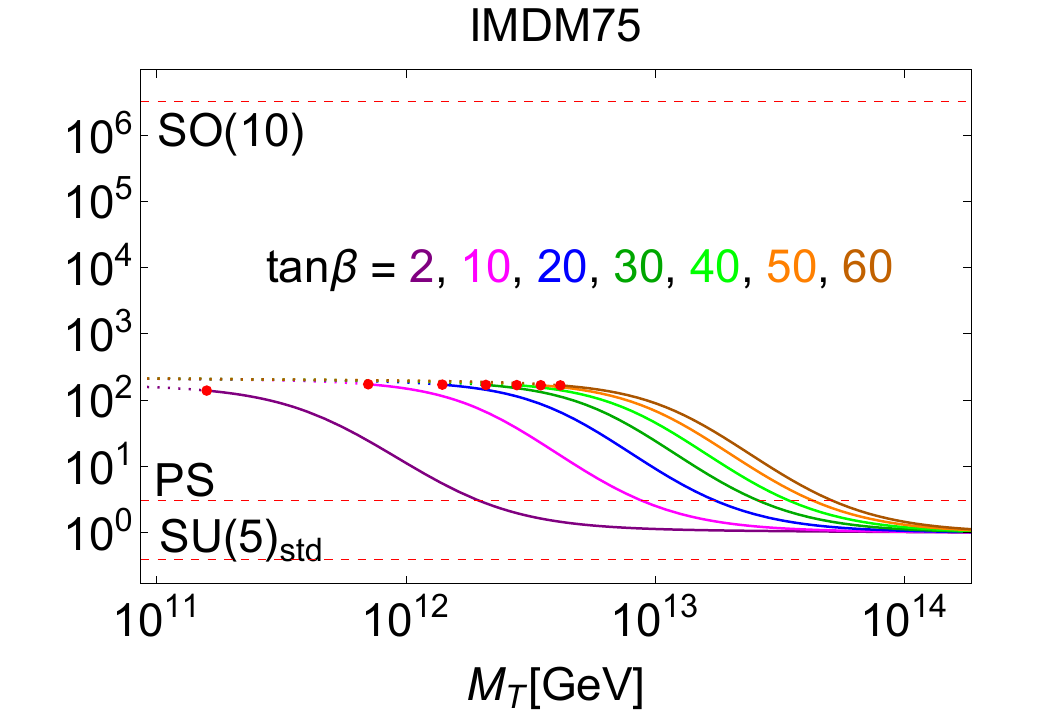}
\includegraphics[width=0.33\textwidth]{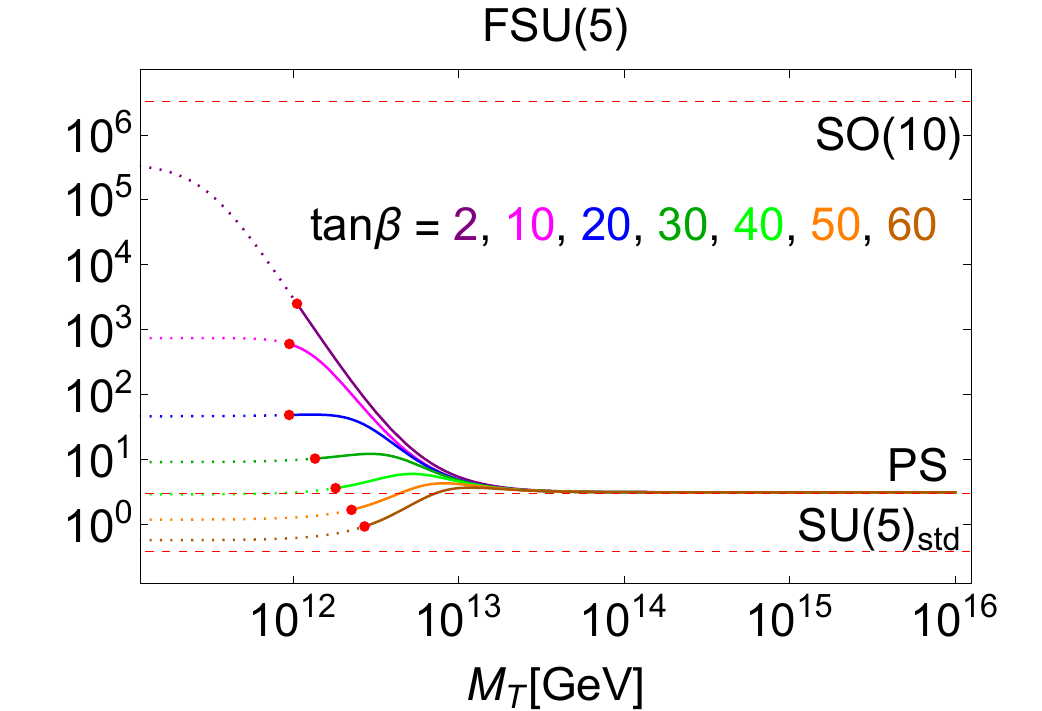}}
\caption{\label{BF1}  The first plot on left side show the estimated values of various branching fractions as a function of $M_T$ in R symmetric $SU(5)$, where $\tan\beta$ varies from $2$ to $60$. In addition to this, the corresponding predicted values of branching fractions for IMDM75 and $FSU(5)$ (middle and right panels), $SU(5)_{std}$, $PS$, and $SO(10)$ (dashed red lines) from \cite{Mehmood:2023gmm, Mehmood:2020irm, Ellis:2020qad, Lazarides:2020bgy, Babu:1997js} are included for comparison. For IMDM75 and $FSU(5)$, $M_T$ refers to the color triplet mass defined in \cite{Mehmood:2023gmm} and \cite{ Mehmood:2020irm} respectively. The solid line curves representing R symmetric $SU(5)$, IMDM75 and $FSU(5)$ predictions are in agreement with the Super-K bounds, as shown by the red dots. Shaded regions on left panels exclude the parameter space according to $\kappa=0.0005$. }
\end{figure*}
Fig.~\ref{nlpd} displays the estimated partial lifetime values vs. $M_T$, for values of  $\tan \beta$ ranging from $2$ to $60$, for the neutral lepton channels. As $M_T$ approaches $10^{14}$~GeV, the gauge boson mediation start dominating over the color-triplet mediation, for most decay channels, and the partial lifetime estimates become independent of $M_T$.

We assume inflaton $\phi$ does not decay into color triplets $(H_T, \bar{H}_T)$ this implies $m_{inf}\leq M_T$, thus proton decay parameter $\lambda = \bar{\lambda}$ is linked with inflation parameter $\kappa$,
\begin{eqnarray}
\kappa &\leq & 
\frac{\sqrt{2} M^3}{M_{50} M_P^2}\lambda ^2.
\end{eqnarray}
This gives us an upper bound on $\kappa$ for $\lambda=1$, variation in the upper bound $\kappa_{max}$ vs. $M$ can be shown in Fig. \ref{kvsM}.
For fixed $M$, $\lambda$ vs. $\kappa$ is given in Fig. \ref{lamvsk}. The minimum value of $\kappa$ set by inflation, puts a lower bound on $\lambda$ and consequently on $M_T$ as well.

Using Super-K  bounds on proton decay channels, lower bounds on the mass of color triplets and involved couplings can be derived as,
\begin{eqnarray}
\lambda 
&\gtrsim &\left(1+\tan^2 \beta\right)^{1/4}\times {\lambda_{min}} \, , \\
 M_T &\gtrsim &\sqrt{1+\tan^2 \beta}\times M_{T}^{min} \, ,
\end{eqnarray}
where $\lambda_{min}$ and  $M_T^{min}$ respectively represents the lower bound on  $\lambda$ and $M_T$  for a given $\tan \beta$ value and decay mode. Similarly, the observable range for $\lambda$  and $M_T$ dependent on $\tan \beta$ can be found using Super-K, Hyper-K and DUNE bounds, as shown in Fig.~\ref{vstb}.
For $ \kappa\geq 0.0005$ lower bound on mass of color triplet is $M_T \geq 3.4\times 10^{13}$ GeV. 
This bound restricts our model for observable proton decay predictions only for $\bar{\nu}_iK^+$ channel for $\tan \beta > 42$. For the rest of the channels our model predict much longer lifetime, that does not lie in the observable range for next generation experiments.

\begin{figure*}[t!]\centering
\subfloat[ \label{mukmupi} $\Gamma_{\mu^+K^0}/\Gamma_{\mu^+\pi^0}$]{\includegraphics[width=0.33\textwidth]{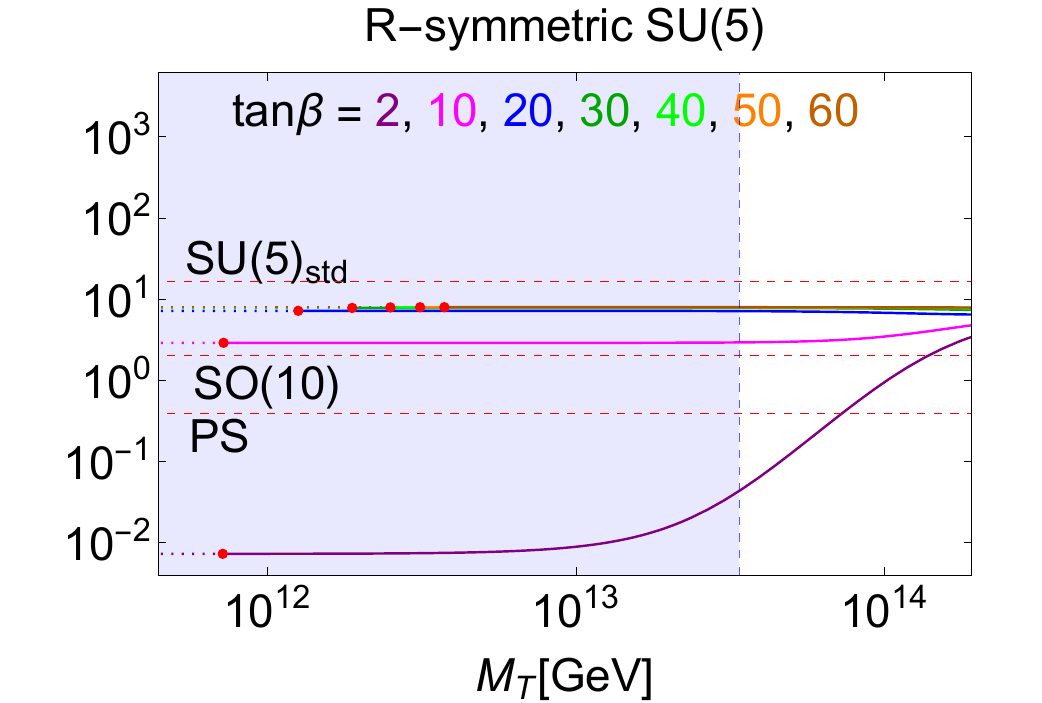}
\includegraphics[width=0.33\textwidth]{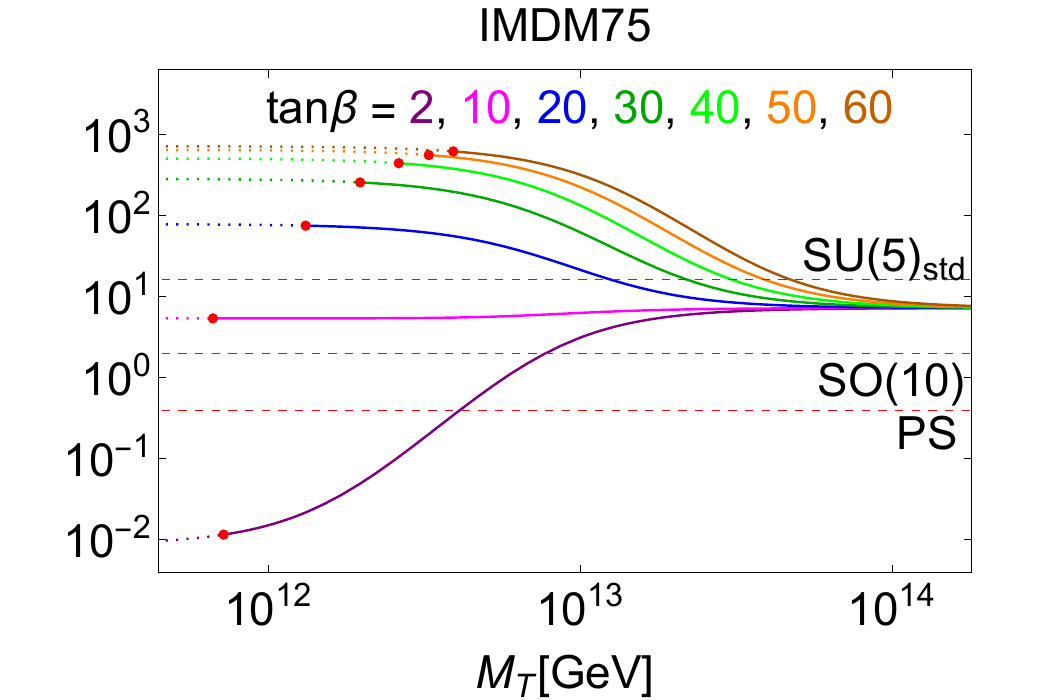}
\includegraphics[width=0.33\textwidth]{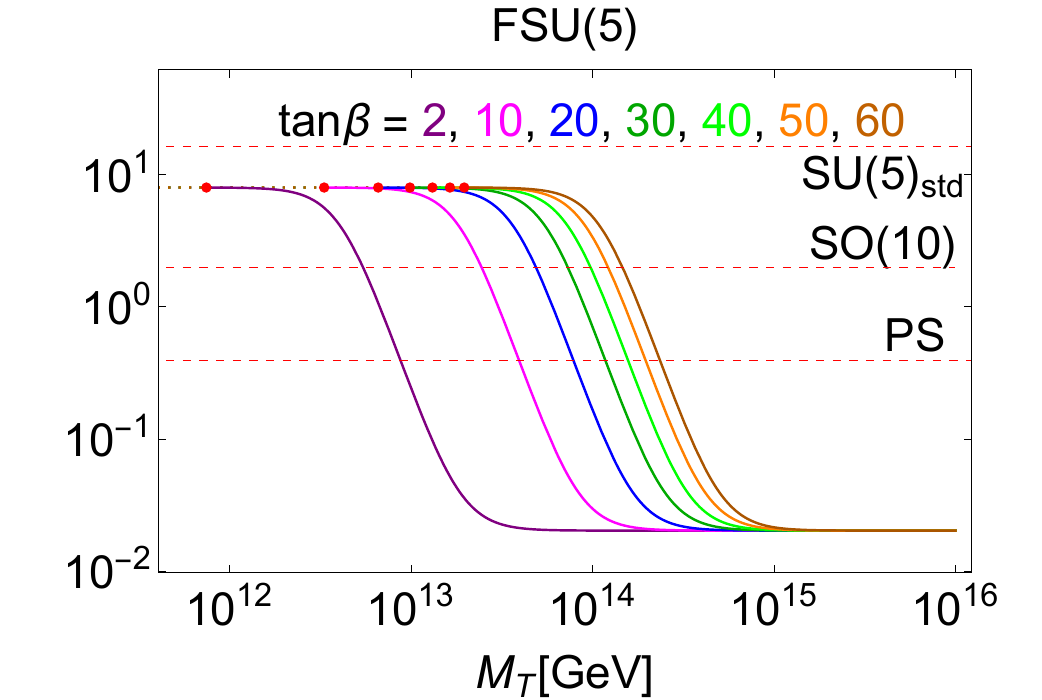}}\\
\subfloat[$\Gamma_{e^+K^0}/\Gamma_{e^+\pi^0}$]{\includegraphics[width=0.33\textwidth]{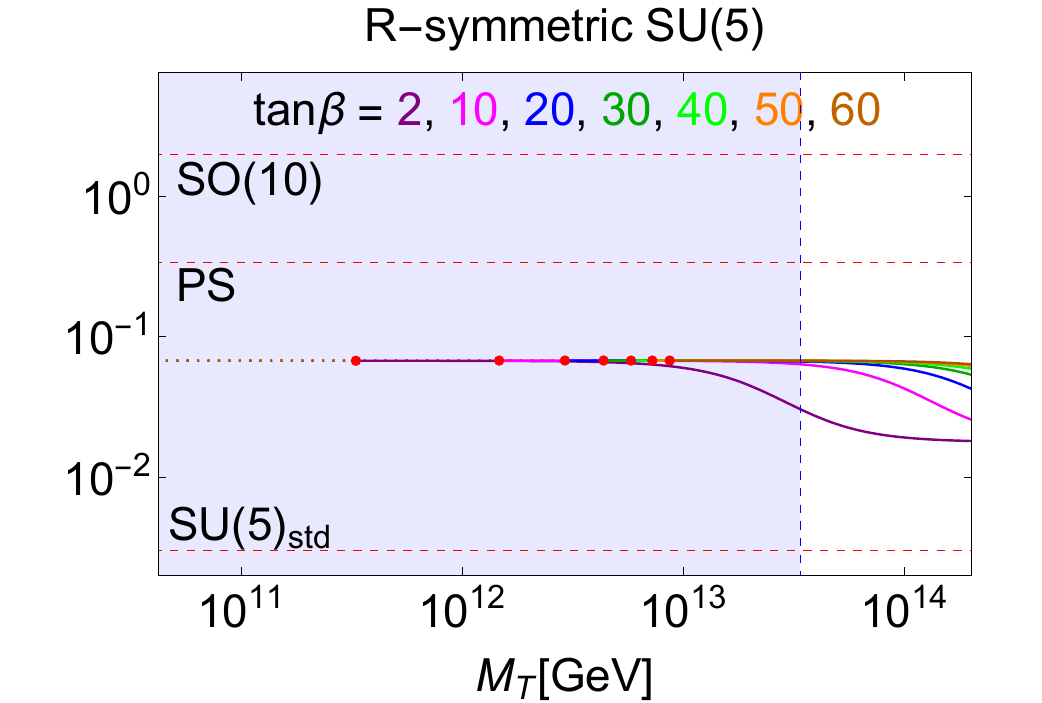}
\includegraphics[width=0.33\textwidth]{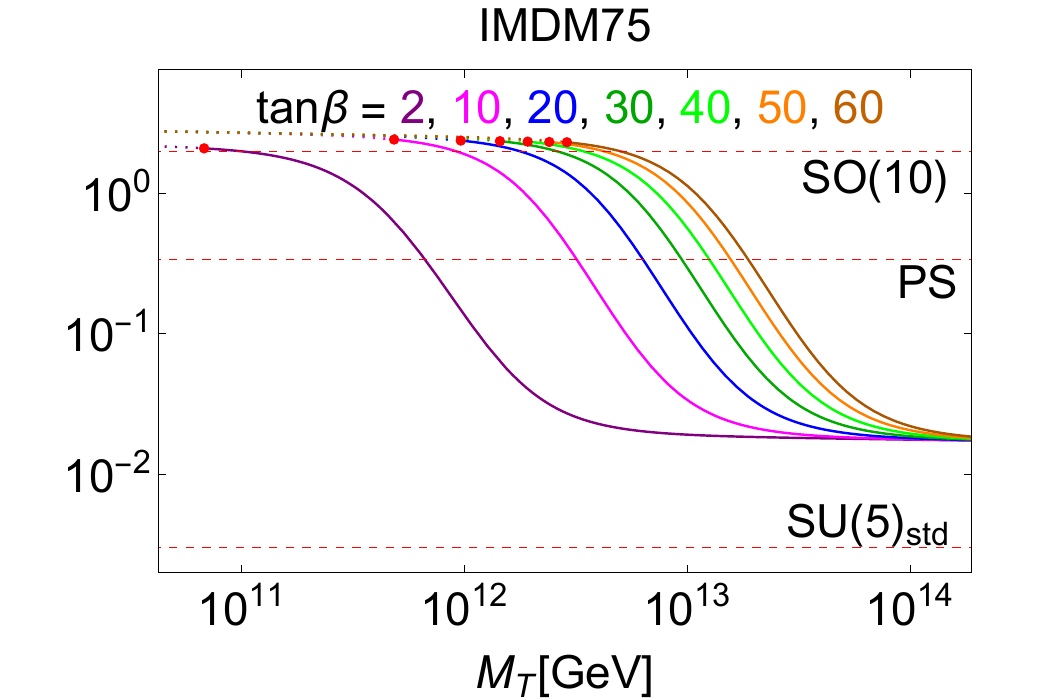}
\includegraphics[width=0.33\textwidth]{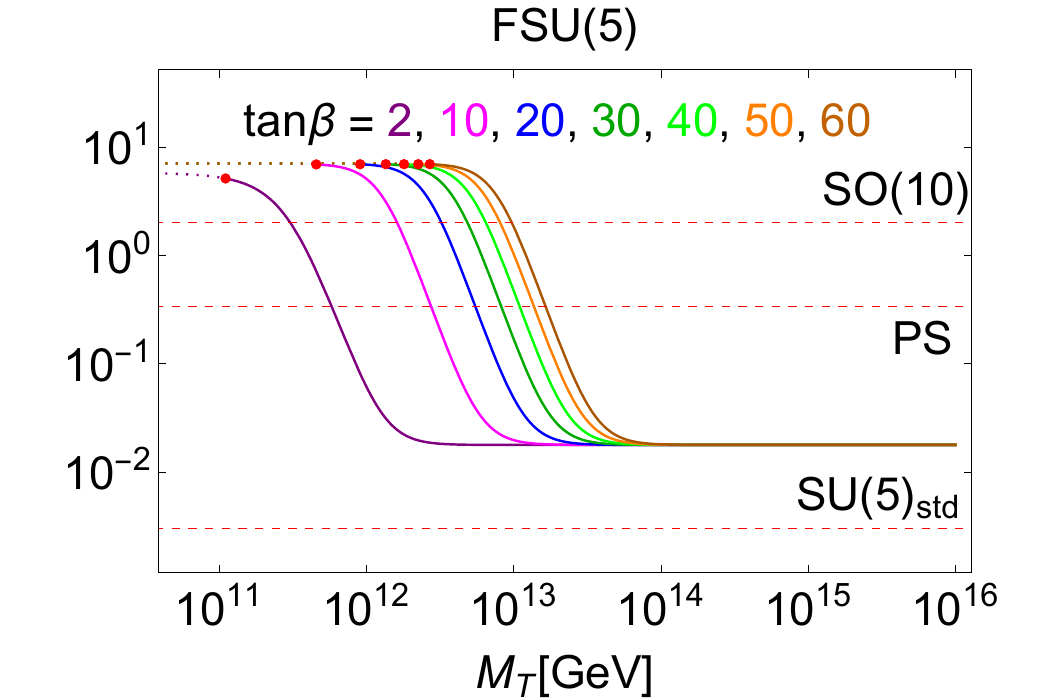}}\\
\subfloat[\label{nuknupi} 
$\Gamma_{\bar{\nu}_iK^+}/\Gamma_{\bar{\nu}_i\pi^+}$]{\includegraphics[width=0.33\textwidth]{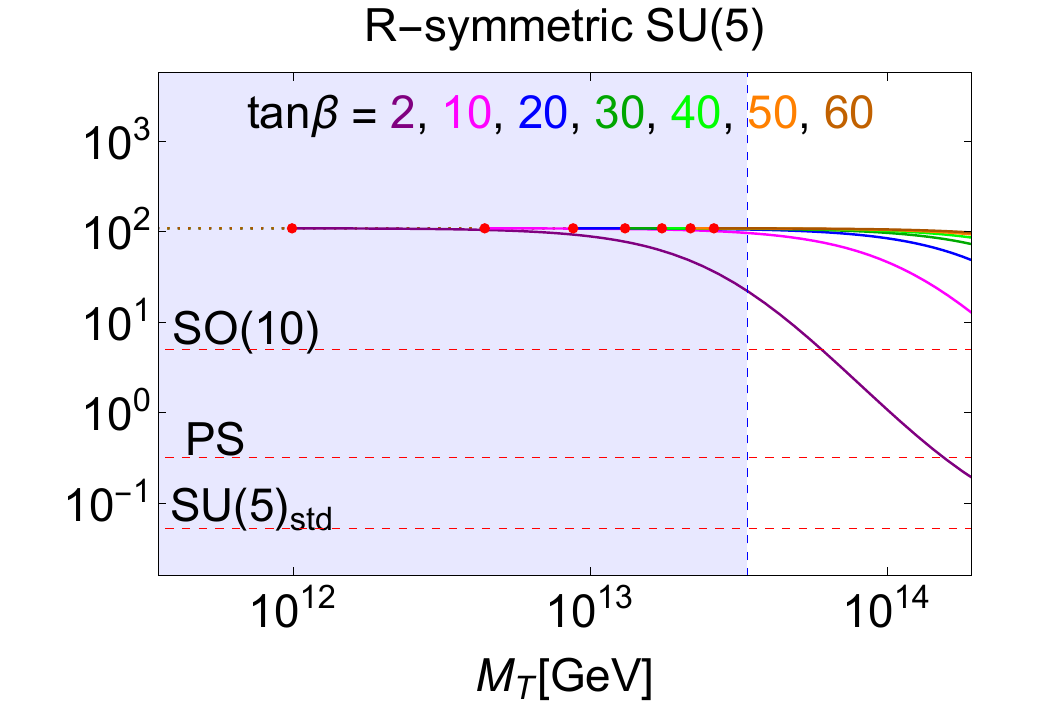}
\includegraphics[width=0.33\textwidth]{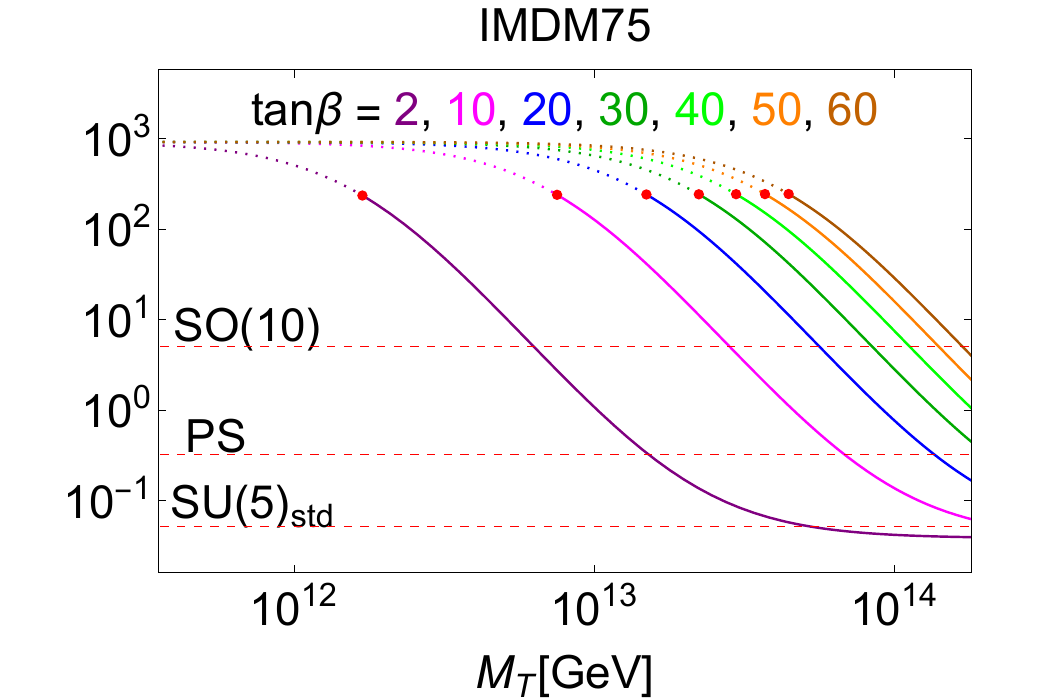} \includegraphics[width=0.33\textwidth]{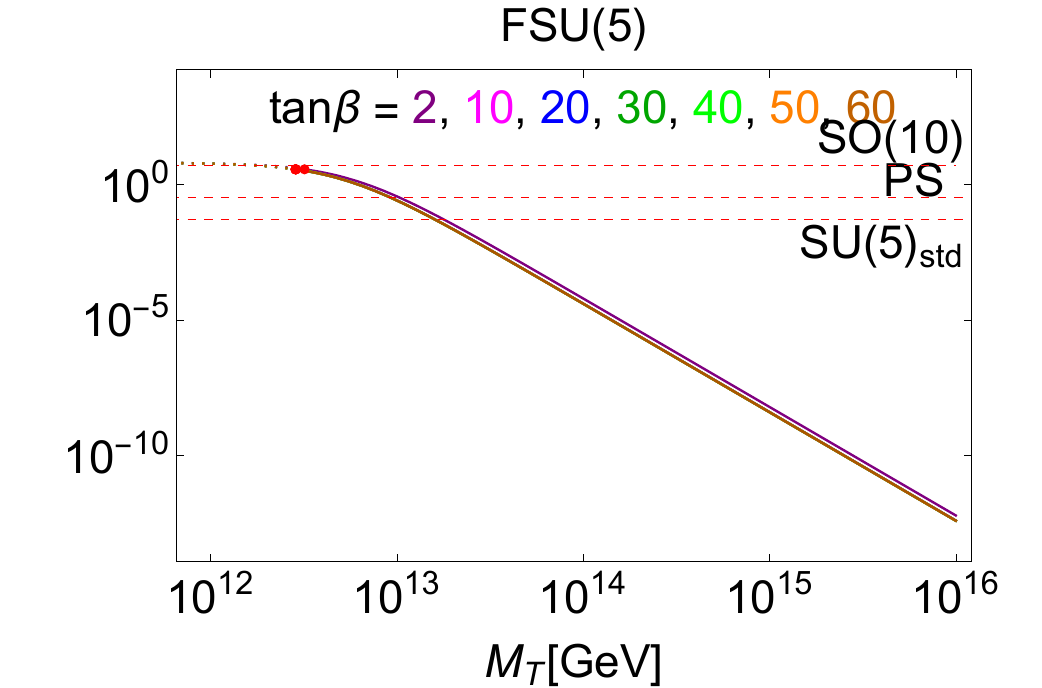}}\\
\subfloat[\label{nukek} $\Gamma_{\bar{\nu}_iK^+}/\Gamma_{e^+K^0}$]{\includegraphics[width=0.33\textwidth]{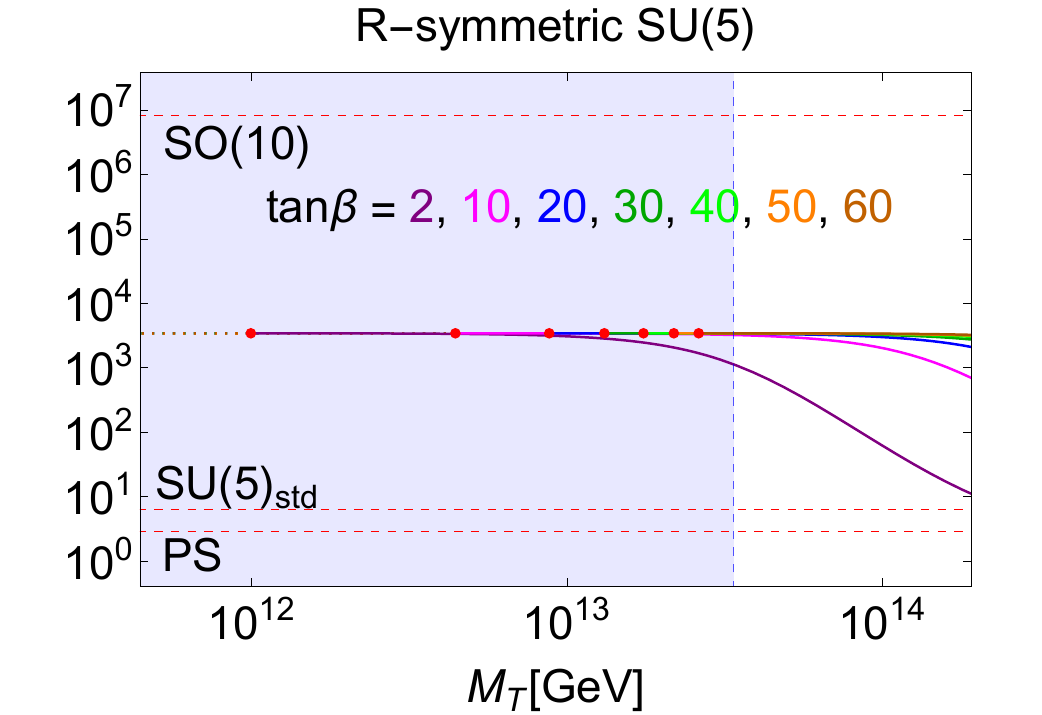}
\includegraphics[width=0.33\textwidth]{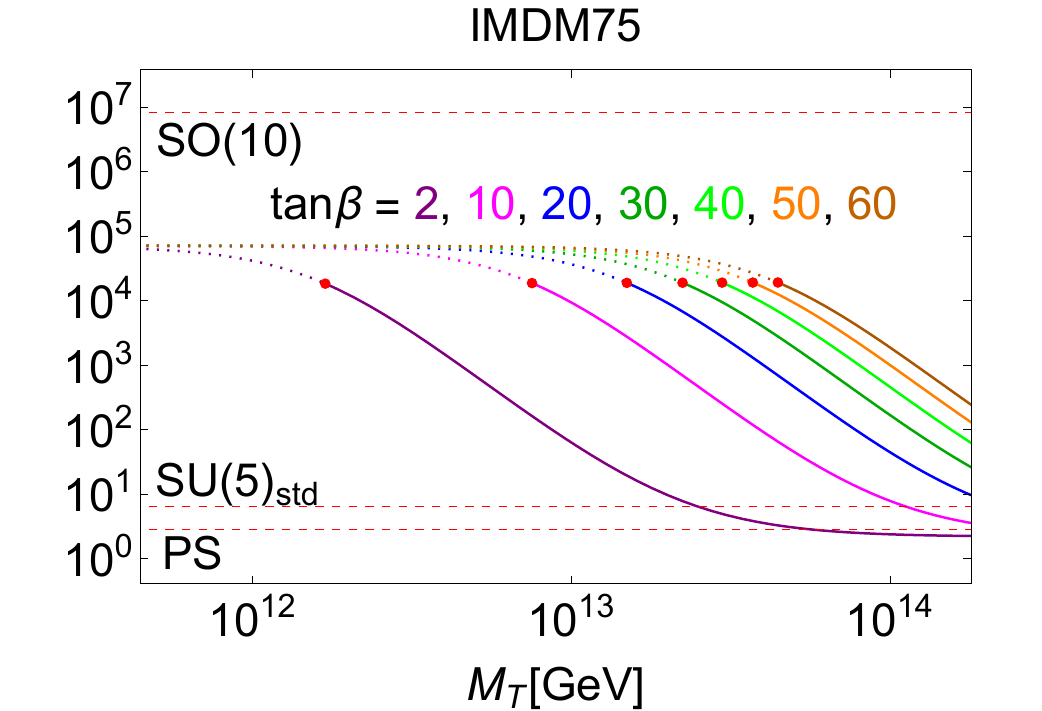}
\includegraphics[width=0.33\textwidth]{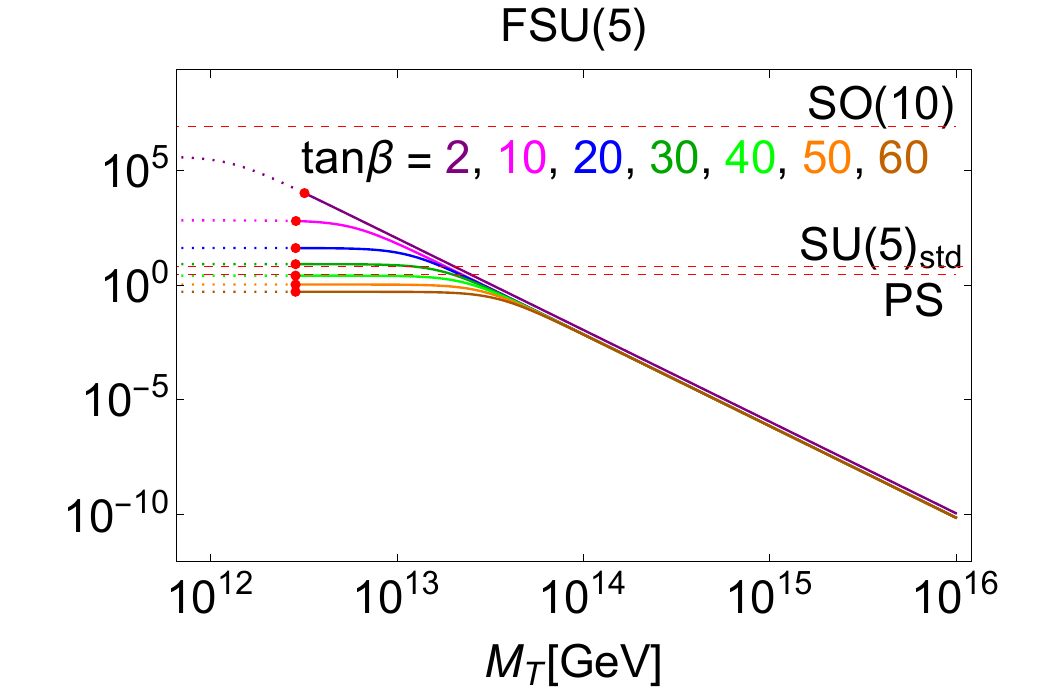}}
\caption{\label{BF2}The first plot on left side show the estimated values of various branching fractions as a function of $M_T$ in R symmetric $SU(5)$, where $\tan\beta$ varies from $2$ to $60$. In addition to this, the corresponding predicted values of branching fractions for IMDM75 and $FSU(5)$ (middle and right panels), $SU(5)_{std}$, $PS$, and $SO(10)$ (dashed red lines) from \cite{Mehmood:2023gmm, Mehmood:2020irm, Ellis:2020qad, Lazarides:2020bgy, Babu:1997js} are included for comparison. For IMDM75 and $FSU(5)$, $M_T$ refers to the color triplet mass defined in \cite{Mehmood:2023gmm} and \cite{ Mehmood:2020irm} respectively. The solid line curves representing R symmetric $SU(5)$, IMDM75 and $FSU(5)$ predictions are in agreement with the Super-K bounds, as shown by the red dots. Shaded regions on left panels exclude the parameter space according to $\kappa=0.0005$. 
}
\end{figure*}
\subsection{R-symmetric $SU(5)$ vs. other models}
To compare the proton decay predictions in R-symmetric $SU(5)$ with other GUT models, we consider the standard minimal $SU(5)$ model with a $24$-dimensional Higgs field \cite{Ellis:2020qad} ($SU(5)_{std}$), nonminimal $SU(5)$ model with a $75$-dimensional Higgs field \cite{Mehmood:2023gmm} (IMDM75), the flipped $SU(5)$ model ($FSU(5)$) \cite{Mehmood:2020irm}, the Pati-Salam model ($PS$) \cite{Lazarides:2020bgy} and the Babu-Pati-Wilczek $SO(10)$ model \cite{Babu:1997js}. Predictions from other $SO(10)$ models can be found in \cite{Babu:1998wi,Haba:2020bls,Djouadi:2022gws}. The branching fractions for various proton decay channels are plotted in Fig.~\ref{BF1} and \ref{BF2}. The first one on left  displays the relationship between the branching fraction and $M_T$ in R symmetric $SU(5)$, while the middle and right side plots presents the corresponding predictions from the IMDM75 and flipped $SU(5)$ model. 

For R symmetric $SU(5)$, shaded regions are excluded by the bound from $\kappa$.
While  R-symmetric $SU(5)$ and IMDM75 places an upper bound on the value of $M_T$, the flipped $SU(5)$ model allows $M_T$ to extend up to $10^{16}$ GeV without any such constraint.

\section{\label{con}Conclusion}
A successful realization of Higgs inflation is discussed by employing a no-scale Kähler potential within a supersymmetric $SU (5)$ GUT with R-symmetry. In this non-minimal Higgs inﬂation scenario, the Higgs ﬁeld $h$ plays the role of the inﬂaton. We set the gauge symmetry breaking scale, $M = 10^{17}$~GeV, and the predicted range of the various inﬂationary observables lie within the $1-\sigma$ bound of Planck 2018 data. The key predictions derived from the effective single field treatment of the model are as follows: the scalar spectral index $n_s\sim0.965$, the tensor-to-scalar ratio $r\sim0.0035$ lies within the testable range of future experiments \cite{Hazumi_2020,Ade_2019,prismcollaboration2013prism,Abazajian_2022,Corbin_2006,sehgal2019cmbhd,Finelli_2018,A_Kogut_2011}, and the running of the spectral index $-dn_s/d\ln k\sim10^{-4}$. A realistic reheating scenario with non-thermal leptogenesis requires the reheat temperature, $T_r\sim10^9$ GeV with the number of e-folds $53.3-54.7$ and the RHN mass in the range, $M_N\sim 10^{11}-10^{12}$. 

The multifield treatment of the model leads to the formation of primordial black holes. Remarkably, within a certain parameter range, these primordial black holes have the potential to entirely account for dark matter. The model also leads to scalar-induced gravitational waves which can be detected by a variety of current and future GW detectors \cite{amaroseoane2017laser,Yagi_2011,Sesana_2021,Sato_2017,Sathyaprakash_2009,Zhao_2013}. The predictions of inflationary observables $n_s$ and $r$ are in good agreement with Planck data. 

A natural realization of doublet-triplet splitting is achieved within the R-symmetric $SU(5)$ model through the utilization of a single pair of $50$-plet. The introduction of R-symmetry breaking at a nonrenormalizable scale facilitated the presence of the color octet and electroweak triplet at an intermediate scale. Consequently, gauge coupling unification is achieved with an octet, electroweak doublet, and color triplets at an intermediate scale.

Proton decay predictions are observable for the decay channel $\bar{\nu}_iK^+$ for $\tan \beta > 42$ and are consistent with cosmological constraint $\kappa \geq 0.0005$. Our model predicts a much longer lifetime of proton for other decay channels that do not lie in the observable range of future experiments. Thus, for next-generation proton decay experiments, $\bar{\nu}_iK^+$ channel will play a distinguishing role in the differentiation of R-symmetric $SU(5)$ with other models.


\section*{Acknowledgments}
N. Ijaz thanks Adeela Afzal for useful discussions related to primordial black holes and gravitational waves.
\FloatBarrier
\bibliography{MDM24}

\end{document}